\documentclass{WileyMSP-template-modified}

\usepackage{cite}
\usepackage[switch,columnwise]{lineno}
\usepackage{lipsum}
\usepackage{threeparttable}
\usepackage{xcolor}
\usepackage{siunitx}
\DeclareSIUnit{\rpm}{rpm}
\DeclareSIUnit{\fps}{fps} 
\usepackage{amsmath}
\usepackage{amssymb}
\usepackage{lettrine}
\usepackage{xfrac}
\usepackage{nicefrac}
\usepackage{bm}
\usepackage{mathtools}
\usepackage[utf8]{inputenc}
\usepackage{booktabs}
\usepackage{arydshln}
\usepackage{array}
\usepackage{stackengine}
\usepackage{ragged2e}
\usepackage{microtype}
\usepackage[]{hyperref}
\hypersetup{
	pdfauthor={Thomas Daunizeau},
	pdfcreator={Thomas Daunizeau},
	pdfproducer={Thomas Daunizeau},
	colorlinks,
	citecolor=black,
	filecolor=black,
	linkcolor=black,
	urlcolor=black
}

\newcommand{\plus}{\raisebox{.3\height}{\scalebox{.7}{+}}}

\justifying

\begin{document}
\pagestyle{fancy}
\rhead{\includegraphics[width=2.5cm]{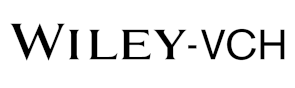}}
\lhead{}

\title{Soft 3D Metamaterial for Low-Frequency Elastic Waves}
\maketitle

\author{Thomas Daunizeau\textsuperscript{*}}
\author{David Gueorguiev}
\author{Vincent Hayward}
\author{Allison Okamura}
\author{Sinan Haliyo}
\dedication{}

\begin{affiliations}
T. Daunizeau, D. Gueorguiev, V. Hayward (deceased), S. Haliyo \par
Sorbonne Université, CNRS, ISIR, F-75005 Paris, France\par
E-mail: thomas.daunizeau@sorbonne-universite.fr\par
A.M. Okamura\par
Department of Mechanical Engineering, Stanford University, Stanford, CA 94305, USA\par
\end{affiliations}


\begin{abstract}
\textbf{Acoustic metamaterials offer exceptional control over wave propagation, but their potential remains unfulfilled due to fabrication constraints. Conventional processes yield mostly rigid, planar structures, whereas soft-matter alternatives have so far been confined to ultrasounds. This work overcomes prior limitations with a fully soft 3D metamaterial operating around \SI{200}{\hertz}. The design combines a 3D-printed elastomer lattice with resonant inclusions of liquid metal, injected via a network of mesofluidic channels. Its dynamic response is derived from a hybrid strategy uniting a lumped-element model with finite element analysis. Simulations~reveal how the dual-phase design decouples flexural and torsional modes, opening a subwavelength band gap for low-frequency elastic waves. Empirical validation is achieved via a custom camera-based vibrometer. Its high spatiotemporal resolution and full-field capabilities enable direct capture of local modes and evanescent waves underlying the band gap. Accelerometer data corroborate these findings and demonstrate greater attenuation than common silicone elastomers at only half of the density. By combining scalable fabrication, compliance, and operations at frequencies relevant to human tactile perception, this novel metamaterial paves the way for lightweight, high-performance cushioning and handles that protect users from harmful vibration exposure.}
\end{abstract}

\section{Introduction}
\label{sec:introduction}

Over the past two decades, acoustic metamaterials have revolutionized wave control~\cite{MaSheng-16, CummerEtAl-16}, enabling effects such as cloaking~\cite{CummerSchurig-07}, wavefront shaping~\cite{XieEtAl-14}, and sub-diffraction focusing~\cite{Pendry-00, LemoultEtAl-11}. By arranging specifically designed unit cells, these engineered media exhibit effective properties, notably frequency-dependent negative mass density $\rho$ and/or modulus $\kappa$~\cite{HuangEtAl-09, LeeEtAl-09, LeeEtAl-10}. This leads to a complex speed of sound~$c$, defined in liquids and solids by $c^2=\kappa/\rho$, which underpins these unprecedented acoustic responses.

A hallmark of acoustic metamaterials is their ability to stop wave propagation in specific frequency ranges, called band gaps. This behavior can stem from Bragg scattering in structures with periodicity commensurate to the wavelength. Alternatively, it may arise from local resonances of subwavelength unit cells, achieving more compact and practical devices. A seminal example is the work of Liu~et~al. on sonic crystals made of rigid metal balls coated in rubber~\cite{LiuEtAl-00}. In essence, this design establishes local gradients in density and rigidity. Beyond its proven performance, this approach stands out for its conceptual simplicity as a practical realization of a lumped mass-spring model. Along these lines, compact subwavelength resonators have been developed by adding extraneous inclusions of materials with high mass density, such as lead~\cite{LiuEtAl-00, OudichEtAl-10} and tungsten~\cite{DuranteauEtAl-16, AstolfiEtAl-21, QuEtAl-22}. Alternatively, the homogeneous bulk can be shaped into thin walls or elongated features~\cite{LeeEtAl-16, DaunizeauEtAl-21}.

Additive manufacturing has been pivotal in unlocking more complex geometries that broaden subwavelength acoustic effects. For instance, 3D-printed spiral metamaterials can make excellent sound diffusers and reflectors~\cite{JimenezEtAl-21, FuEtAl-18}, render holograms~\cite{XieEtAl-16}, and create mufflers from Fano-like interferences~\cite{GhaffarivardavaghEtAl-19}. While early metamaterials were mostly planar, 3D printing has promoted~volumetric designs, such as Luneburg lenses~\cite{XieEtAl-18} and truss elements in body-centered cubic lattices~\cite{AnEtAl-20}. However, current 3D-printing technologies only induce moderate impedance variations within the volume. Such shortfall is evident when comparing a 3D-printed lattice of period \SI{50}{\milli\meter}, with a band gap starting at about \SI{2}{\kilo\hertz}~\cite{AnEtAl-20}, to a conventional approach that confers a band gap starting at \SI{400}{\hertz}, despite being only \SI{15.5}{\milli\meter} in size\cite{LiuEtAl-00}. Multi-material printing is a promising solution, but limited to polymers whose molecular compositions yield similar densities, around $\SI[inter-unit-product = \!\cdot\!]{1.1}{\g\per\centi\meter\cubed}$.

In parallel, recent breakthroughs in soft matter printing have enabled highly compliant, stretchable 3D lattices~\cite{TrubyLewis-16, Skylar-ScottEtAl-19}, spurring the development of a wide spectrum of mechanical metamaterials~\cite{BabaeeEtAl-13, BertoldiEtAl-17, JiangWang-16}. To date, however, these structures are limited to quasi-static operations despite the vast potential of soft 3D printing for dynamic wave control. Instead, existing soft acoustic metamaterials are made using alternative methods such as air-filled porous silicones~\cite{LeroyEtAl-15, BaEtAl-17, JinEtAl-19}, liquid-filled gels~\cite{ZhangEtAl-19a}, or randomly distributed colloids suspended in gels~\cite{BrunetEtAl-13, BrunetEtAl-15, ZhangEtAl-24}. These suspensions can be arranged periodically with optical~\cite{BaumgartlEtAl-07} or acoustic tweezers~\cite{CaleapDrinkwater-14}, in which case they exhibit Bragg band gaps. Partially soft acoustic metamaterials have also been reported, with Minnaert resonances from air bubbles trapped in a nylon scaffold~\cite{CaiEtAl-19} or resonant microbeads entwined in soft fibrous networks~\cite{TangEtAl-18}. Overall, these non-printing approaches favor small, lightweight inclusions only suited to ultrasonic regimes. Access to sub-\SI{}{\kilo\hertz} frequencies remains an open problem, hindering deployment in haptic devices and seismic isolators, which involve large wavelengths typically exceeding \SI{10}{\centi\meter} and \SI{100}{\meter}, respectively.

\begin{figure}[!b]
\centering
\captionsetup{width=178mm}
\includegraphics[width=178mm]{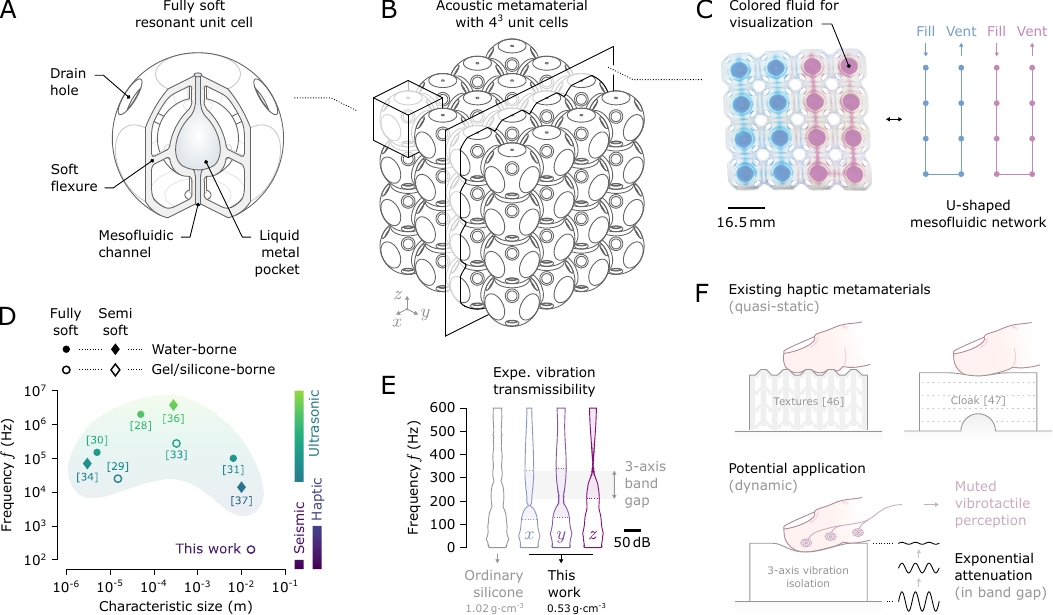}
    \caption{A) Cross-section of a unit cell with a resonant inclusion of liquid metal held by orthogonal soft rods. B) Schematic of our fully soft 3D metamaterial, restricted to a cubic lattice of $4^3$ unit cells for practicality. C) Slice revealing the \mbox{U-shaped} mesofluidic channels used for filling. Colored fluid is for visualization only. D) Comparison with state-of-the-art soft metamaterials for elastic or acoustic wave propagation across aqueous, silicone, and gel-like media. Our design operates at frequencies nearly two orders of magnitude lower. Characteristic size is defined as the lattice constant for periodic structures or as the microparticle diameter for disordered colloidal dispersions. E) Transmissibility measurements along the three axes of our metamaterial. Narrower violin plots indicate stronger attenuation. Band gaps are shown as shaded areas, with their overlap highlighted. Our metamaterial achieves greater attenuation than ordinary silicone, here taken as Ecoflex 00-10, at a fraction of the density. F) Potential use in haptics as a low-frequency vibration insulator, contextualized alongside existing quasi-static haptic metamaterials.}
\label{fig:overview}
\end{figure}

We address these challenges by introducing a novel soft 3D metamaterial for low-frequency elastic wave control. Inspired by advances in liquid metal transducers~\cite{JungYang-15, WangEtAl-19}, our design features a compliant cubic lattice with resonant inclusions of low-melting-point alloy. We employ a hybrid analytical and numerical framework to efficiently open a band gap, circumventing the high computational cost of topology optimization~\cite{LiEtAl-19a}, transformation acoustics~\cite{PendryEtAl-06}, or brute-force finite element analyses (FEA). We further propose a robust, fully soft prototyping process combining 3D-printing with~mesofluidics. For experimental validation, our custom vibrometry method builds upon~\cite{WadhwaEtAl-17} to provide detailed, direct observation of local resonances and evanescent waves associated with the band gap. Finally, we benchmark subwavelength properties and attenuation against standard rubber substrates.

\vspace{-2mm}
\section{Results and Discussion}
\label{sec:results}

\subsection{Design Rationale}
\label{sec:design}

Designing a subwavelength resonant unit cell requires the reconciliation of two conflicting goals: reducing its size while lowering its resonance frequency. This demands compact and compliant~structures, which can be achieved through a multi-scale approach involving both the intrinsic flexibility of rubbery materials and thin macroscopic features. Soft stereolithography (SLA) printing was chosen for manufacturing as it is ideally suited to produce such characteristics. A soft resin was preferred for its low elastic modulus of about \SI{1.8}{\mega\pascal} (see Experimental Section). An orthogonal arrangement of \SI{1.0}{\milli\meter} diameter soft rods was designed to create an isotropic flexure, as shown in \textbf{Figure~\ref{fig:overview}}.A. It was encased within an outer shell, resulting in a monolithic unit cell that is self-supported, a crucial requirement for SLA printing. Drain holes were added at each corner to remove uncured resin. As depicted in Figure~\ref{fig:overview}.B, these unit cells were arranged in a cubic lattice, chosen for its practicality.

The resulting compliant structure lacked the necessary mass to make a resonator. Once 3D printed, its volume becomes difficult to access, hindering the usual method of inserting rigid metallic inclusions. Instead, we created lumped masses at the intersections of the soft rods~by~post-print injection of a dense fluid into spherical pockets (see Figure~\ref{fig:overview}.A and Movie~S1, Supporting Information). To~achieve sufficient density, metals that are liquid at room temperature are promising candidates. We selected Galinstan as a non-toxic alternative to mercury ($\SI[inter-unit-product = \cdot]{13.5}{\gram\per\centi\meter\cubed}$), while offering a density~of~$\SI[inter-unit-product = \cdot]{6.44}{\gram\per\centi\meter\cubed}$\! at \SI{20.0}{\celsius}. Galinstan is a eutectic alloy of gallium (68.5~wt\%), indium (21.5~wt\%), and tin (10.0~wt\%), with a melting point of \SI{10.7}{\celsius}.

A primary design challenge was ensuring the distribution of Galinstan to each unit cell. This~was achieved by linking the spherical pockets via a network of mesofluidic channels. Each channel~extended to the outermost layer of the metamaterial, providing access for both filling and venting ports. Several topologies were explored with channels of \SI{0.9}{\milli\meter} inner diameter (see Section~S1.1~and~Figure~S1, Supporting Information). These channels were routed through the soft rods, whose outer diameters were adjusted accordingly. Balancing fabrication complexity while reducing pressure build-up during fluid injection required dividing the network into U-shaped channels, as shown in Figure~\ref{fig:overview}.C.

Our dual-phase design can be tuned (see Section~\ref{sec:optimization}) to operate at frequencies up to two~orders of magnitude lower than existing soft acoustic metamaterials while remaining compact and subwavelength (see Figure~\ref{fig:overview}.D). We specifically target the \SI{200}{\hertz} range, relevant to human touch~\cite{JohanssonEtAl-82, BolanowskiEtAl-88}. This supports dynamic effects, unlike existing lattices aimed solely at quasi-static interactions, such as patterning textures~\cite{IonEtAl-18} or concealing objects from touch~\cite{BuckmannEtAl-14} (see Figure~\ref{fig:overview}.F). By stopping elastic waves, our design could advance localized vibrotactile feedback, improving on rigid 1D devices~\cite{DaunizeauEtAl-25}, or damp tool-induced vibrations to mitigate hand–arm vibration syndrome (HAVS)~\cite{DongEtAl-21}.

\begin{figure}[!t]
\centering
\includegraphics[width=178mm]{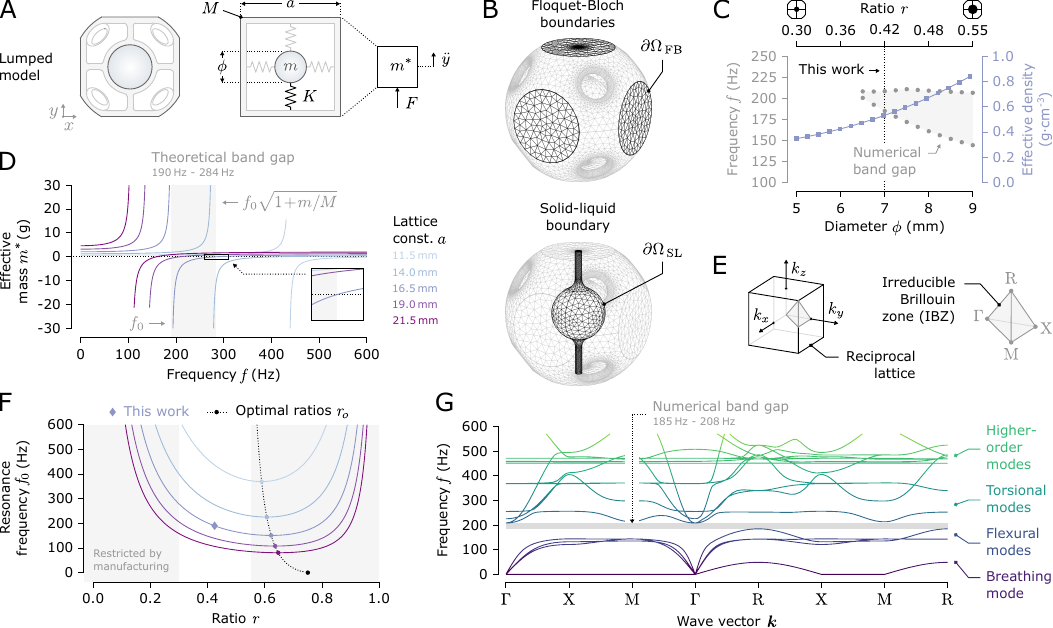}
\caption{A) Lumped 1D mass-spring model of a unit cell. B) Meshed 3D finite-element model of a unit cell. Specific boundaries are highlighted. C) Numerically computed band gap and density for $a=\SI{16.5}{\milli\meter}$. Markers are discrete simulation steps. The dotted line shows the chosen design. D) Frequency-dependent effective lumped mass for lattice constants $a$ and a ratio $r=0.42$. The shaded region outlines a preliminary band gap estimate. E) Tetrahedral irreducible Brillouin zone. F) Resonance frequency of the lumped model as a function of $a$ and $r$. Shaded areas denote $r$ values excluded by fabrication constraints. The dotted curve shows the theoretical optimum $r_{\!o}$, from which the chosen design deviates to ensure manufacturability. G) Dispersion graph of the tuned lattice with $a=\SI{16.5}{\milli\meter}$ and $r=0.42$. A complete band gap is bounded by flexural and torsional modes.}
\label{fig:optimization}
\end{figure}

\subsection{Theoretical and Numerical Optimization}
\label{sec:optimization}

Achieving the strong attenuation and low density summarized in Figure~\ref{fig:overview}.E (see details in Section~\ref{sec:attenuation}) requires finely-tuned dimensions. We first approximated the unit cell by a lumped-element model to rapidly identify promising designs. As shown in \textbf{Figure~\ref{fig:optimization}}.A, the liquid metal pocket of diameter $\phi$ was treated as a rigid lump of equal mass $m$, and the flexure was modeled by an effective stiffness $K$, derived from Euler-Bernoulli beam theory (see Figure~S2 and Section~S1.3, Supporting Information). This subsystem of resonance frequency $f_0$ was nested inside an outer shell of mass $M$. Assuming isotropy and neglecting mesofluidic channels, the motion was restricted to one dimension. Under dynamic equilibrium, the unit cell is equivalent to a homogeneous medium with an effective mass $m^*\!$, negative for frequencies from $f_0$ to $f_0\sqrt{1\!+\!m/M}$ (see Figure~\ref{fig:optimization}.D and Section~S1.2, Supporting Information). This range defines a band gap where the acceleration $\Ddot{y}$ counteracts the external force $F$, thus stopping wave propagation. Both band gap limits scale with $f_0$ and can be expressed in terms of two design variables to optimize: the lattice constant $a$ and the dimensionless ratio $r=\phi/a \in]0,1[$, reflecting the relative size of the liquid inclusion. Accordingly, Figure~\ref{fig:optimization}.F provides guidance on choosing $(a,r)$ to coarsely position the band gap. For example, a band gap near \SI{400}{\hertz}, as in rigid sonic crystals~\cite{LiuEtAl-00}, could be achieved with a $26\%$ smaller lattice constant of \SI{11.5}{\milli\meter}. We chose $a=\SI{16.5}{\milli\meter}$ to target a band gap around \SI{200}{\hertz}, valid across a broad range of $r$. There exists an optimal ratio $r_{\!o}$ that minimizes $f_0$ and, in turn, yields the most subwavelength unit cell. However, for $a=\SI{16.5}{\milli\meter}$, $r_{\!o}=0.62$ was unattainable due to soft SLA limits (e.g. minimum wall thickness of \SI{1}{\milli\meter}) which restricted $r$ to $[0.30,0.55]$. These constraints and lumped simplifying assumptions motivated a refined analysis to determine $r$ and precisely tune the band gap.

Building on this, we developed a detailed 3D model to capture intricate geometries, mesofluidic channels, solid-liquid coupling, nonlinear material laws, and higher-order dynamics (see Section~S2, Supporting Information). Assuming infinite 3D periodicity, the metamaterial was reduced to a single unit cell with periodic Floquet-Bloch boundary conditions, as shown in Figure~\ref{fig:optimization}.B. Soft matter was treated as hyperelastic to account for finite strains and the liquid metal followed a Newtonian fluid. With no closed-form solution, this eigenproblem was solved via FEA (see Experimental Section and Table~S1, Supporting Information), made tractable by prior reduction of the design space. Results from numerical iterations over $r$ are reported in Figure~\ref{fig:optimization}.C. A band gap opens for $r>0.39$ ($\phi=\SI{6.5}{\milli\meter}$). Its lower limit decreases with $r$ whereas its upper limit remains mostly unchanged. We selected $r=0.42$ ($\phi=\SI{7.0}{\milli\meter}$) to set a band gap from \SI{185}{\hertz} to \SI{208}{\hertz}, balancing the need for a sufficiently wide gap and a lightweight structure. Its lower limit matches theory ($+6\%$) whereas the upper limit deviates slightly more ($-21\%$), likely due to torsional modes omitted by the lumped model. The dispersion diagram in Figure~\ref{fig:optimization}.G reveals a complete band gap for any wave vector $\boldsymbol{k}$ along the irreducible Brillouin zone (IBZ), as defined in Figure~\ref{fig:optimization}.E. This confirms the quasi-isotropic nature of the 3D-printed flexure, with limited impact from the mesofluidic channels. This is key for handling elastic waves at any incidence angle, as in the tactile applications previously discussed.

\begin{figure}[!b]
\vspace{-2mm}
\centering
\includegraphics[width=178mm]{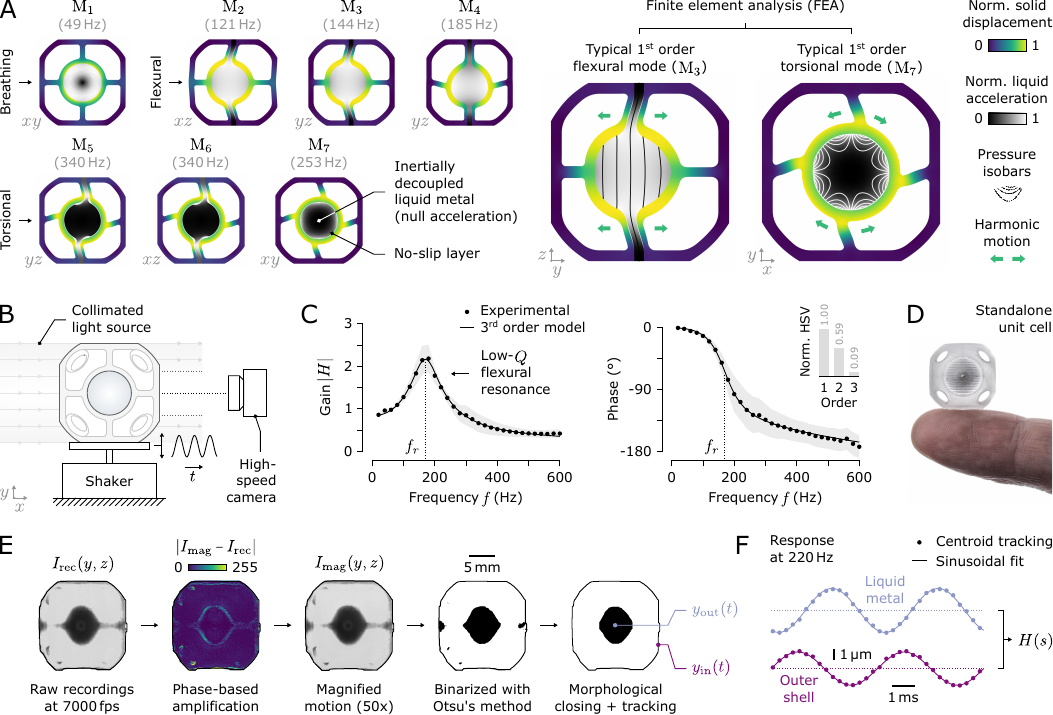}
\caption{A) Finite element analysis of local vibration modes framing the band gap in an infinite lattice. The deformed shapes, magnified for clarity, are depicted at wave vector $\boldsymbol{k}^\mathrm{X}$\! for $\mathrm{M}_2$ and $\mathrm{M}_3$, with $\boldsymbol{k}^\mathrm{R}$\! for the remaining modes. Normalized amplitudes of solid displacement and liquid acceleration are shown, with pressure isobars overlaid. B) High-speed imaging setup for full-field vibrometry. C) Bode diagram, averaged over ten standalone unit cells and fitted by a third-order model. The shaded area shows $\pm 1$ SD. Inset shows the Hankel singular value decomposition. D) Photograph of a standalone unit~cell, about the size of a fingertip. E)~Image-processing workflow for tracking the vertical motion of the liquid metal and outer shell centroids. F) Resulting empirical harmonic response, e.g. for a \SI{220}{\hertz} input of standalone cell No.~3.}
\label{fig:local_resonances}
\end{figure}

\vspace{-2mm}
\subsection{Taxonomy of Local Resonances}
\label{sec:local_resonances}

A fully soft unit cell prompts questions about how solid-liquid coupling shapes the band gap. The branches in Figure~\ref{fig:optimization}.G outline complex dynamics beyond that of a simple lumped model. The lowest branch corresponds to mode $\mathrm{M}_1$ depicted in \textbf{Figure~\ref{fig:local_resonances}}.A, a monopolar breathing of the fluid with spatially uniform pressure, constrained by the elasticity of the spherical pocket. Mode $\mathrm{M}_1$ occurs at frequencies low enough to leave the band gap unaffected. In contrast, branches framing the gap arise from first-order flexural and torsional modes (see Figure~\ref{fig:local_resonances}.A and Movie~S2, Supporting Information). 

Using FEA, we identified three flexural modes, $\mathrm{M}_2, \mathrm{M}_3, \mathrm{and}\, \mathrm{M}_4$, corresponding to bending along the Cartesian basis $\{\boldsymbol{e}_i \,|\, i \in \! \{x,y,z\}\}$. As illustrated in Figure~\ref{fig:local_resonances}.A, these modes deflect the soft rods and induce a pressure gradient within the fluid, resulting in a slightly pear-shaped pocket. In turn, the liquid metal undergoes a near-uniform acceleration along the direction of motion. At the critical point~X, the wave vector $\boldsymbol{k}^\mathrm{X}=\pi/a\,[0 1 0]$ yields pure bending, enabling direct comparison with the lumped model. For a wave vector $\boldsymbol{k}^\mathrm{R}=\pi/a\,[1 1 1]$, $\mathrm{M}_2$ and $\mathrm{M}_3$ become degenerate at \SI{143}{\hertz} (see Figure~S3.B, Supporting Information). Mesofluidic channels enlarge and stiffen the soft rods they pass through, raising $\mathrm{M}_4$ frequency to \SI{185}{\hertz}, 29\% higher than that of degenerate modes $\mathrm{M}_2$/$\mathrm{M}_3$. This shift was small enough to maintain the band gap, whose lower limit was thus set by $\mathrm{M}_4$.

The branches directly above the gap are dominated by torsional modes $\mathrm{M}_5, \mathrm{M}_6, \mathrm{and}\, \mathrm{M}_7$. The pair $\mathrm{M}_5$/$\mathrm{M}_6$ becomes degenerate at the critical points R (\SI{340}{\hertz}) and X (\SI{369}{\hertz}). As with flexion, the mesofluidic channels introduce anisotropy, making $\mathrm{M}_7$ frequency 26\% smaller than that of $\mathrm{M}_5$/$\mathrm{M}_6$ for~$\boldsymbol{k}^\mathrm{R}$. Consequently, the upper limit of the band gap is set by the branch containing $\mathrm{M}_5$ and whose minimum occurs at either $\Gamma$ or M, depending on the ratio~$r$ (see Figure~S3.A, Supporting Information). As noted in Figure~\ref{fig:local_resonances}.A, a no-slip boundary layer forms along the inner wall of the spherical pocket, shearing the fluid and generating local pressure gradients. Owing to the low dynamic viscosity of Galinstan, $\mu=\SI[inter-unit-product = \!\cdot\!]{2.4}{\milli\pascal\second}$, about twice that of water, shear stays confined to this layer and leaves the bulk stagnant, as shown by the dark-shaded regions in Figure~\ref{fig:local_resonances}.A. Unlike flexion, torsion thus engages only a negligible fraction of the liquid metal inertia. Instead, replacing the liquid with a solid inclusion of equal density would involve most of the inertia, regardless of the vibration mode (see Figures~S4, Supporting Information). This hypothetical case would lead to overlapping branches and close any gap. An overlap may also occur due to poorly chosen dimensions, for instance when $r<0.39$, as seen in Figure~\ref{fig:optimization}.C. Collectively, these findings indicate that a strong contrast between flexion and torsion is key to open a low-frequency band gap. Such modal differentiation was readily achieved using a high-density, low-viscosity fluid, and was mostly unaffected by mesofluidic channels, thereby validating our design.

The experimental visualization of vibration modes with FEA-like level of detail proves challenging. However, our fully soft design deforms sufficiently to enable standard cameras to capture flexural modes (see Movie~S3, Supporting Information). Ten standalone unit cells were prototyped with dimensions $a=\SI{16.5}{\milli\meter}$ and $r=0.42$ (see Experimental Section and Figure~\ref{fig:local_resonances}.D). Their translucent shell exposed the otherwise concealed internal resonator, which prompted the development of a full-field optical vibrometry setup using high-speed imaging (see Experimental Section). As illustrated in Figure~\ref{fig:local_resonances}.B, harmonic excitation at frequencies $f\! \in \! \{20,40,\dots,600\}\,\SI{}{\hertz}$ was applied by a shaker along $\boldsymbol{e}_y$, thus driving primarily $\mathrm{M}_3$. Similar findings are expected for $\mathrm{M}_2$ and $\mathrm{M}_4$. As shown in Figure~\ref{fig:local_resonances}.E, we extracted the output~$y_\mathrm{out}(t)$ and input~$y_\mathrm{in}(t)$, taken as the liquid metal and outer shell centroids, respectively. As shown in Figure~\ref{fig:local_resonances}.F, they both fit sinusoids ($\forall \, f,\, \mathrm{RMSE}\leq\SI{0.19}{\micro\meter}$), consistent with the forced harmonic response of a time-invariant system.

We derived the transfer function $H(s)\!=\!Y_\mathrm{out}(s)/Y_\mathrm{in}(s)$, with $s$ the Laplace variable. As shown by the Bode plot in Figure~\ref{fig:local_resonances}.C, the data are in excellent agreement with a third-order state-space model~($\mathrm{RMSE}=0.03$, see Section~S3.1, Supporting Information). Low variability in $|H|$~across all unit cells ($\forall \, f,\,\mathrm{SD}\leq 0.34$) indicates reproducible fabrication and measurement. The Hankel singular~value (HSV) decomposition in the inset of Figure~\ref{fig:local_resonances}.C reveals two high-energy states and a third, weaker state, presumably reflecting fluid rheology or strain-rate effects. This is supported~by~pole analysis, which gives a real pole and a complex-conjugate pair (see Figure~S5.B, Supporting Information). The~resulting undamped natural frequency, $f_n=\SI{176}{\hertz}$, closely matches that of the corresponding FEA stand\-alone mode $\mathrm{M}_3$ (\SI{177}{\hertz}) with less than 1\% error (see Figure~S5.C, Supporting Information). A~minor discrepancy was expected as the mass of Galinstan per unit cell ($\SI{1.057}{\gram}$) averages $9\%$ below theory ($\SI{1.157}{\gram}$), likely due to reduced fill volume from 3D-printing distortions. Damping lowers the resonance to $f_r=\SI{170}{\hertz}$ and imparts a low quality factor $Q=1.81$. Our setup effectively revealed the first flexural mode, though torsion remained out of reach. As shown in Figure~\ref{fig:local_resonances}.C, at resonance, the liquid metal oscillates $\pi/2$ out of phase with the shaker. This creates an inertial force opposed to motion, offering a straightforward Newtonian explanation for how flexion may initiate a band gap.

\begin{figure}[!b]
\centering
\includegraphics[width=178mm]{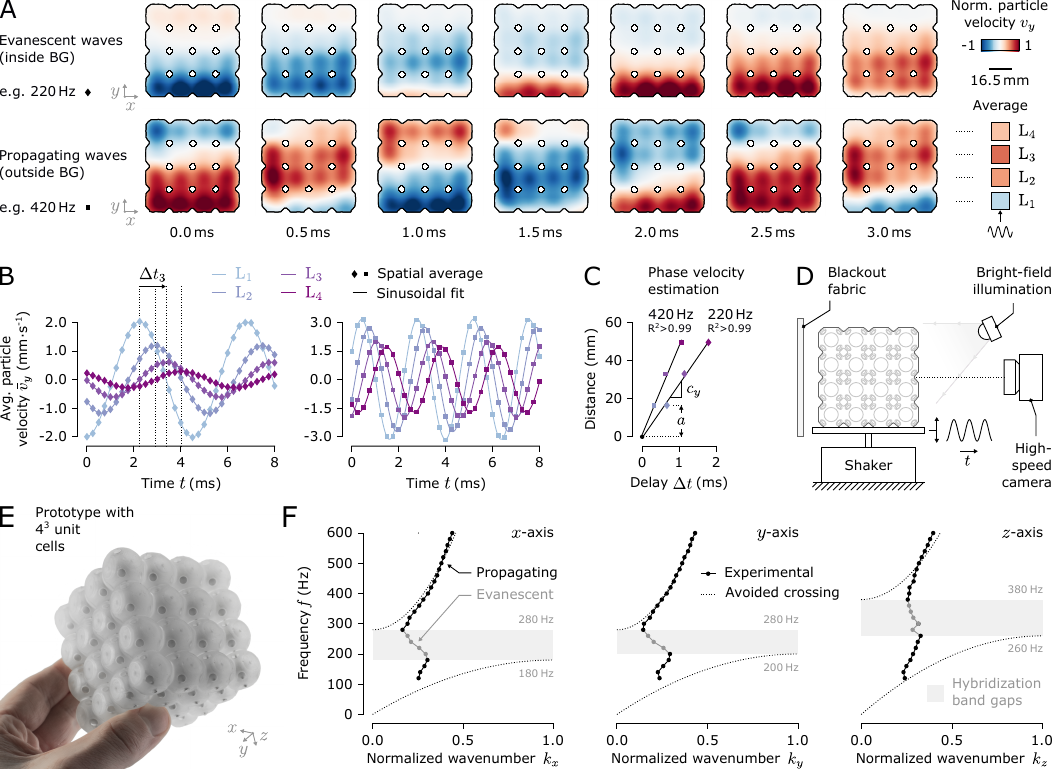}
\caption{A) Spatiotemporal evolution of the $y$-axis particle velocity shown at \SI{0.5}{\milli\second} intervals, i.e. one in every two frames. Examples are given for \SI{220}{\hertz} and \SI{420}{\hertz} sine inputs, inside and above the band gap $\mathrm{BG}_y$, respectively. The top row reveals evanescent waves while the bottom row shows waves propagating throughout the metamaterial. B) Spatially averaged $y$-axis particle velocity in each layer $\mathrm{L}_j$ for \SI{220}{\hertz} and \SI{420}{\hertz}. C) Phase velocity derived by linear regression of the phase lag between layers. D) Optical vibrometry setup for capturing waves on the metamaterial outer shell. E) Photograph of the prototype made of $4^3$ unit cells. F) Empirical dispersion diagrams. Shaded areas denote hybridization band gaps. Dotted curves are second-order polynomial model of the propagating modes, fitted to the data with tangency enforced at the band gap boundaries.}
\label{fig:band_gap}
\end{figure}

\subsection{Band Gap Properties}
\label{sec:band_gap}

A comprehensive evaluation revealed low-$Q$ flexural resonances in single unit cells. To verify if they could collectively form a subwavelength band gap, we fabricated a metamaterial with $4^3$ unit cells, as shown in \textbf{Figure~\ref{fig:band_gap}}.E (see Experimental Section). Practical considerations limited its size. As depicted in Figure~\ref{fig:band_gap}.D, the optical setup was adapted to record waves on the outer shell of the metamaterial rather than exposing its inner structure (see Experimental Section).

To evaluate 3D properties, measurements were repeated by successively aligning each principal axis $\{\boldsymbol{e}_i \,|\, i \in \! \{x,y,z\}\}$ of the metamaterial with the vertical input. The particle velocity field~$\boldsymbol{v}$~was computed by optical flow, as illustrated in Figure~\ref{fig:band_gap}.A for \SI{220}{\hertz} and \SI{420}{\hertz}. The wavefronts are slightly tilted, likely due to uneven actuation. For each layer of unit cells $\{\mathrm{L}_j \,|\, j \in \! [1,4]\}$, the spatially averaged particle velocity $\bar{\boldsymbol{v}}$ aligned closely with a sinusoid ($\forall \, (f,\boldsymbol{e}_i,\mathrm{L}_{j}), \, \mathrm{RMSE}\leq\SI[inter-unit-product = \!\cdot\!]{0.08}{\milli\meter\per\second}$), as shown in Figure~\ref{fig:band_gap}.B. This confirms the accuracy of an optical reconstruction of the wavefield. The wavefronts passed through successive layers with a time delay $\Delta t_j$ ($j \geq 2$) found via cross\nobreakdash-correlation. Since layers are equally spaced by the lattice constant $a$, the phase velocity $\boldsymbol{c}$ was found by linear regression ($\forall \, (f,\boldsymbol{e}_i), \, \mathrm{R}^2 \geq 0.97$), as shown in Figure~\ref{fig:band_gap}.C and Figure~S6, Supporting Information. In~turn, the wave vector $\boldsymbol{k}=2\pi f/\boldsymbol{c}$ yields the dispersion diagrams in Figure~\ref{fig:band_gap}.F.

Data provided evidence of an avoided crossing, characteristic of a resonant metamaterial. For each axis $\boldsymbol{e}_i$, the resulting hybridization band gap $\mathrm{BG}_i$ was identified from the evanescent segment. Conversely, propagating bands were approximated by simple quadratics ($\forall \, \boldsymbol{e}_i, \mathrm{RMSE}\leq\SI{74}{\hertz}$), chosen for parsimony, with horizontal tangents enforced at $\|\boldsymbol{k}\|\!=\!\{0,\pi/a\}$. Outliers below \SI{120}{\hertz} were excluded. Residuals likely arise from finite-size effects, since a four-layer prototype violates the infinite lattice assumption. While $\mathrm{BG}_x$ and $\mathrm{BG}_y$ nearly coincide, $\mathrm{BG}_z$ moves to higher frequencies, mirroring the upward frequency shift of mode $\mathrm{M}_4$ induced by the mesofluidic channels (see Section~\ref{sec:local_resonances}). For excitation along $\boldsymbol{e}_y$, Figure~\ref{fig:band_gap}.A shows that elastic waves outside $\mathrm{BG}_y$ (e.g.~at~\SI{420}{\hertz}) propagate across the metamaterial, whereas waves within $\mathrm{BG}_y$ (e.g.~at~\SI{220}{\hertz}) are evanescent (see Movie~S4, Supporting Information). At peak attenuation, i.e. \SI{240}{\hertz} for waves along~$\boldsymbol{e}_y$ (see Section~\ref{sec:attenuation}), the wavelength $\lambda_y$ is \SI{164}{\milli\meter}, indicating a deeply subwavelength response with a dimensionless ratio $\lambda_y/a=9.9$ (see Section~S3.3 for other axes, Supporting Information). Overall, despite its low quality factor, our fully soft design successfully opens a 3-axis band gap. This wave-based analysis complements Section~\ref{sec:local_resonances} and further emphasizes the role of flexion in opening the band gap.

\vspace*{-2mm}
\subsection{Global Attenuation Performance}
\label{sec:attenuation}

\begin{figure}[!t]
\centering
\includegraphics[width=178mm]{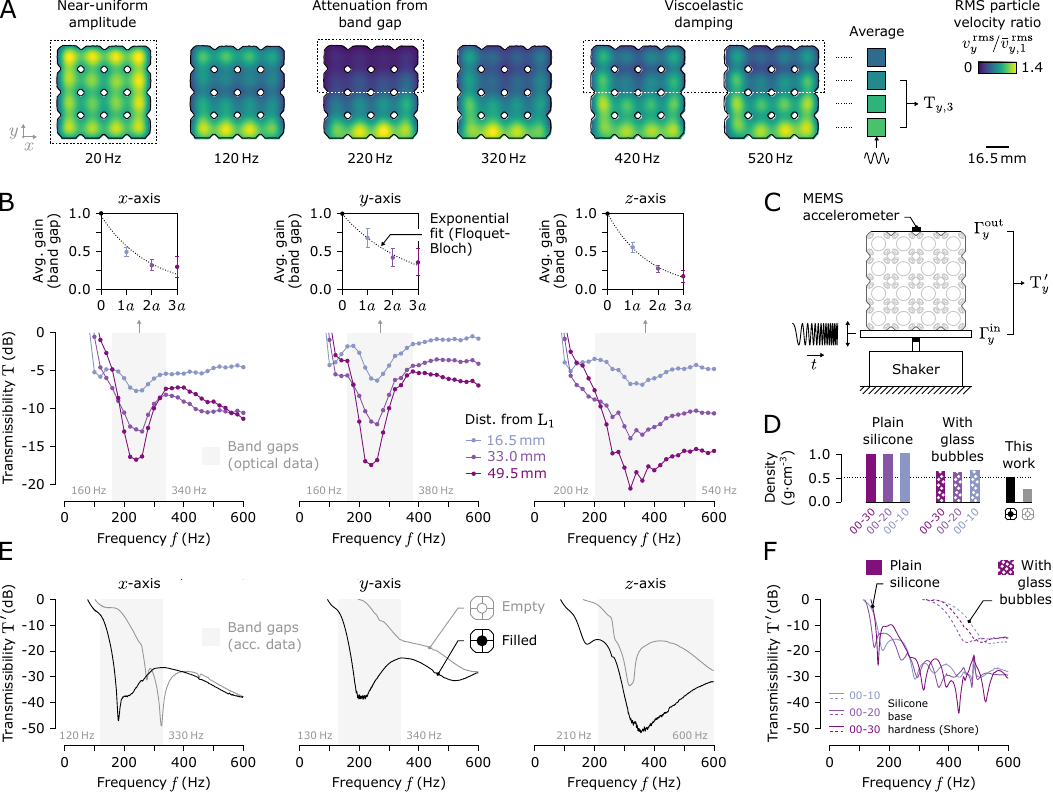}
\caption{A) RMS particle velocity along the $y$-axis, normalized to layer $\mathrm{L}_1$ input. Distinct frequency regimes exhibit \mbox{near-uniform} motion, strong band gap attenuation, and weaker viscoelastic damping. B) Transmissibility $\mathrm{T}$ from optical~vibrometry, along each axis and for each layer $\mathrm{L}_j$. Band gaps are shaded, with insets showing gap-averaged gains, fitted to an exponential decay as in Floquet-Bloch theory. C) Experimental setup for global transmission measurements using accelerometers. D) Density of the metamaterial and common silicones, either plain or mixed with micro glass bubbles. E) Transmissibility $\mathrm{T}^{\,\prime}$\! from accelerometer data for the metamaterial either empty or filled, averaged over ten trials. F) Corresponding results for the reference silicones.}
\label{fig:attenuation}
\end{figure}

A band gap $\mathrm{BG}_i$ was observed on each axis $\boldsymbol{e}_i$, but the attenuation it provided has yet to be quantified. To address this, we combined optical vibrometry with conventional accelerometers. The RMS particle velocity field $\boldsymbol{v}^\mathrm{rms}$, derived from prior optical recordings, is shown in \textbf{Figure~\ref{fig:attenuation}}.A for waves along $\boldsymbol{e}_y$ (see Figures~S7~to~S9 for the full dataset, Supporting Information). The transmissibility was defined as $\mathrm{T}_{i,j} = 20\log\,(\bar{v}^\mathrm{\,rms}_{i,j}/\bar{v}^\mathrm{\,rms}_{i,1})$ with $\bar{v}^\mathrm{\,rms}_{i,j}$ the RMS particle velocity along $\boldsymbol{e}_i$, spatially averaged within each layer $\mathrm{L}_j$. We only report data for $T \leq 0\,\SI{}{\deci\bel}$ since bulk resonance lies beyond this study. Figure~\ref{fig:attenuation}.B shows significant attenuation in each band gap $\mathrm{BG}_i$ with $\min\,\{\mathrm{T}_{x,4}, \mathrm{T}_{y,4}, \mathrm{T}_{z,4}\}=-\{17, 17, 21\}\,\SI{}{\deci\bel}$ at $\{240, 240, 320\}\,\SI{}{\hertz}$, respectively. The similar spectral responses along $\boldsymbol{e}_x$ and $\boldsymbol{e}_y$ contrast with the persistent high-frequency attenuation along $\boldsymbol{e}_z$, corroborating the modal anisotropy caused by the mesofluidic channels (see Section~\ref{sec:local_resonances}). Without a universally accepted empirical definition~of~band gap limits, we defined them as the inflection points in Figure~\ref{fig:attenuation}.B. The resulting limits align~well with findings in Figure~\ref{fig:band_gap}.F, though they yield slightly wider gaps. The insets of Figure~\ref{fig:attenuation}.B indicate that the transmission coefficients, $\bar{v}^\mathrm{\,rms}_{i,j}/\bar{v}^\mathrm{\,rms}_{i,1}$, averaged across the band gap, decay exponentially ($\forall \, \boldsymbol{e}_i, \, \mathrm{RMSE}\leq 0.1$), as predicted by Floquet-Bloch theory.

While optical vibrometry offers superior spatial resolution for surface recordings, accelerometers enable transmission analysis across the volume with finer frequency steps (see Experimental Section). An alternative transmissibility was thus defined as $\mathrm{T}_{i}^{\,\prime} = 20\log(\lvert\Gamma_{\!i}^\mathrm{out}\rvert/\lvert\Gamma_{\!i}^\mathrm{in}\rvert)$, with $\Gamma_{\!i}^\mathrm{out}$\!~and~$\Gamma_{\!i}^\mathrm{in}$ the Fourier transforms of normal accelerations on opposite faces of the metamaterial, as illustrated in Figure~\ref{fig:attenuation}.C. Pre-injection measurements established a baseline reflecting the viscoelastic damping of the soft SLA resin, as shown in~Figure~\ref{fig:attenuation}.E. Along $\boldsymbol{e}_y$, adding liquid metal opened a band~gap, outlining its pivotal role in enabling local resonances. Interestingly, we observed pre-existing attenuation notches along $\boldsymbol{e}_x$ and $\boldsymbol{e}_z$. Given the wavelengths involved, they could not arise from Bragg scattering and are more consistent with antiresonances of the silicone lattice. Liquid metal then broadened and shifted these notches. Band gaps found using accelerometers, rounded to the nearest \SI{10}{\hertz}, closely matched those derived optically (see Table~S2 for a summary, Supporting Information). Therefore, not only the optically accessible outermost unit cells, but also those within the bulk, operate as designed, validating our 3D architecture with U-shaped mesofluidic channels. The complete band gap, defined as $\mathrm{BG}_x \cap \mathrm{BG}_y \cap \mathrm{BG}_z$, spans $[200, 340]\,\SI{}{\hertz}$ for velocity data and $[210, 330]\,\SI{}{\hertz}$ for acceleration data. Although on a similar scale, both empirical ranges are broader than the numerical prediction of $[185, 208]\,\SI{}{\hertz}$. This likely stems from a combination of prototyping irregularities, viscoelasticity, and finite-size effects, all of which favor broadband characteristics over sharp attenuation.

Despite similar trends, $\mathrm{T}_{i}^{\,\prime}$\! curves are \SI{30}{\deci\bel} lower than $\mathrm{T}_{i,4}$, likely due to weak bonding of the top accelerometer, which limited energy transfer. This did not affect benchmarking against off-the-shelf elastomers, which relied solely on $\mathrm{T}_{i}^{\,\prime}$. Prior tests were repeated on cubic silicone samples of matching size, spanning a range of hardness and densities achieved using different formulations, with or without micro glass bubble fillers (see Figure~\ref{fig:attenuation}.D and Experimental Section). Comparing Figures~\ref{fig:attenuation}.E~and~\ref{fig:attenuation}.F shows our metamaterial outperformed other options with $\min\,\{\mathrm{T}_{x}^{\,\prime}, \mathrm{T}_{y}^{\,\prime}, \mathrm{T}_{z}^{\,\prime}\}=-\{47, 38, 52\}\,\SI{}{\deci\bel}$ at $\{181, 210, 356\}\,\SI{}{\hertz}$, respectively, while~having an effective density of only $\SI[inter-unit-product = \cdot]{0.53}{\gram\per\centi\meter\cubed}$. Although hollow fillers are often used to enhance damping via scattering from air pockets, they proved counterproductive at low frequencies. This further highlights the benefit of dense liquid resonators. As summarized in Figure~\ref{fig:overview}.E, our soft-matter approach delivers strong attenuation at half the density of rubber, enabling lightweight, tunable, low-frequency~3D~isolation.

\newpage
\section{Conclusion}
\label{sec:conclusion}

We reported a novel soft metamaterial made of liquid metal resonators embedded within a compliant, 3D-printed elastomer lattice. To our knowledge, this is the first fully soft system to exhibit band gaps for low-frequency elastic waves (typically below \SI{1}{\kilo\hertz}). The dense, low-viscosity fluid decouples flexural and torsional modes, opening low-frequency band gaps. This suggests a broader principle in which combining materials in distinct physical states can prevent modal superposition and, in turn, opens band gaps via avoided crossing. Liquid metal was injected into otherwise inaccessible volumes through a network of mesofluidic channels, streamlining fabrication and eliminating labor-intensive manual assembly of rigid inclusions. By selectively adjusting the size of each liquid metal pocket, this paves the way for hierarchical and tunable unit cells. Such tailoring potential may provide practical solutions for transformation acoustics, supporting novel volumetric lenses and cloaks.

To navigate the geometric complexity, large deformations, nonlinearities, and solid-liquid coupling inherent to our fully soft system, we introduced a hybrid workflow pairing a lumped-element analytical model with finite-element analysis. This approach enabled efficient optimization over a two-variable design space with reduced computational cost.

Experimental validation was conducted by prototyping and testing both individual unit cells and a $4^3$-cell metamaterial. Leveraging high-speed imaging, we developed a full-field vibrometry~system capturing flexural modes wherein the liquid metal oscillated out of phase with the lattice. Despite a low intrinsic quality factor, these soft matter resonators collectively induced a complete hybridization band gap across $[200, 340]\,\SI{}{\hertz}$. Optical vibrometry was substantiated by accelerometer data showing a $[210, 330]\,\SI{}{\hertz}$ band gap. We further confirmed these results by optically mapping evanescent waves in unprecedented detail, revealing the exponential decay predicted by Floquet-Bloch theory.~With only four layers of unit cells, our metamaterial provides strong attenuation of up to \SI{21}{\deci\bel} (optical~estimate). At an effective density of $\SI[inter-unit-product = \!\cdot\!]{0.53}{\gram\per\centi\meter\cubed}$, it outperforms common elastomers for a fraction of the mass. Its compact \SI{16.5}{\milli\meter} cubic unit cells ensure deep-subwavelength operation, stopping propagation at wavelengths about ten times their size.

This work lays a comprehensive foundation for soft metamaterials operating at frequencies nearly two orders of magnitude lower than current alternatives. With further miniaturization, we foresee impactful haptic applications, for instance as padding or handle grips that damp harmful vibrations. In this context, future work will focus on validating user benefits through psychophysical experiments that measure tactile perceptual thresholds as a proxy for effective vibration attenuation. Conversely, reaching even lower frequencies, such as in seismic regimes, will require exponentially larger structures since local resonances vary as $1/a^2$, which our scalable contribution readily supports.

\newpage
\section{Experimental Section}
\label{sec:experimental}

\threesubsection{\hspace*{1.2em}Numerical Simulations}\\
The eigenvalue problem was solved using finite element analysis in COMSOL Multiphysics 6.2. As shown in Figure~\ref{fig:optimization}.B, a single unit cell was modeled with periodic Floquet-Bloch boundary conditions on each face, $\partial\Omega_{\,\mathrm{FB}}$, of its cubic envelope. Mesh elements on opposing faces were mirrored to satisfy Floquet-Bloch requirements. Solid-liquid coupling was implemented via a no-slip condition at the interface, $\partial\Omega_{\,\mathrm{SL}}$, while the remaining surfaces were left free. The model was meshed with over $10^5$ quadratic tetrahedral elements, with further refinement in regions of high-curvature. Solid domains followed a nearly incompressible Neo-Hookean hyperelastic model with a Poisson's ratio $\nu=0.48$, a density $\rho_s=\SI[inter-unit-product = \cdot]{1.02}{\gram\per\centi\meter\cubed}$, and Young's modulus $E=\SI{1.8}{\mega\pascal}$ derived from tensile testing data given by the resin manufacturer. Solid elements satisfied the wave equation under finite strain theory, accounting for geometric nonlinearity due to large deformations. Liquid domains were modeled as a Newtonian fluid with a dynamic viscosity $\mu=\SI[inter-unit-product = \!\cdot\!]{2.4}{\milli\pascal\second}$, a phase velocity $c_l=\SI[inter-unit-product = \!\cdot\!]{2730}{\meter\per\second}$\!, and a density $\rho_l=\SI[inter-unit-product = \!\cdot\!]{6.44}{\gram\per\centi\meter\cubed}$. Liquid elements satisfied the linearized Navier-Stokes equation.

\threesubsection{Prototyping}\\
The $4^3$ lattice was 3D-printed in $\SI{100}{\micro\meter}$-thick layers via stereolithography (Form 3\plus, Elastic 50A V1, Formlabs) as described in Figure~S10, Supporting Information. This process left uncured resin in the spherical pockets and mesofluidic channels, which required a multi-step removal (see Figure~S11, Supporting Information). These channels were first vacuumed with a syringe, then flushed with isopropyl alcohol (IPA). These steps were repeated until all residual resin was removed. To mitigate swelling caused by IPA immersion, the part was dried in an oven at \SI{80}{\celsius} for \SI{2}{\hour}. Post-curing with UV was omitted to maintain a low modulus of about \SI{1.8}{\mega\pascal}. One extremity of each channel was sealed by a droplet of resin, hardened using a \SI{5}{\milli\watt}, \SI{405}{\nano\meter}-UV laser for \SI{2}{\second}. This resulted in U-shaped mesofluidic channels which facilitated the injection of liquid metal while preventing the formation of air pockets. The remaining open ends were sealed to complete the fabrication. The prototype was filled with \SI{71.1}{\gram} of Galinstan and weighed a total \SI{151.2}{\gram}, for an effective static density of \SI[inter-unit-product = \!\cdot\!]{0.53}{\gram\per\centi\meter\cubed}. In addition, ten standalone unit cells were made following the same process. On average, they were filled with $\SI[separate-uncertainty, multi-part-units=single]{1.057 \pm 0.009}{\gram}$ (mean $\pm$ SD) of Galinstan and weighed a total of $\SI[separate-uncertainty, multi-part-units=single]{2.058 \pm 0.010}{\gram}$, for an effective density of $\SI[separate-uncertainty, multi-part-units=single,inter-unit-product = \!\cdot\!]{0.458 \pm 0.002}{\gram\per\centi\meter\cubed}$ (see Table~S4, Supporting Information).

\threesubsection{Silicone Samples}\\
Cubic silicone samples of side \SI{66}{\milli\meter} were cast to match the metamaterial dimensions. Three plain silicone samples with Shore hardness of 00-10, 00-20, and 00-30 (Ecoflex Series, Smooth-On) were prepared. Each two-part silicone was mixed in a 1:1 mass ratio, vacuum degassed for \SI{5}{\minute} at \SI{6}{\kilo\pascal}, poured into 3D-printed polylactic acid (PLA) molds, and cured at \SI{20}{\celsius} for \SI{4}{\hour}. An additional set of three samples was prepared by adding micro glass bubble filler (K20, 3M, \SI{60}{\micro\meter} average diameter) at a 1:8 mass ratio. Corresponding properties are listed in Table~S3, Supporting Information.

\threesubsection{Vibrating Apparatus}\\
The custom vibrating stage, designed to suppress undesired resonances from \SI{0}{\hertz} to \SI{600}{\hertz}, was built using high-modulus carbon fiber plates (see Figures~S12.A and~S12.B, Supporting Information). Longitudinal waves were generated by an electrodynamic shaker (2185, \SI{8}{\ohm}, Frederiksen) and transmitted via a moving plunger. To prevent high-frequency detachment, the standalone unit cells and the metamaterial were secured to the vibrating stage using either thin adhesive tabs (\SI{9}{\milli\meter} PELCO Tabs 16084-3, Ted Pella Inc.) or pressure-sensitive tape (9087, 3M). Driving signals were sampled at $\SI{20}{\kilo\hertz}$ by a 16-bit acquisition card (PCI-6221, National Instrument) and fed to a class-D amplifier with a \SI{20}{\deci\bel} gain (TPA3112D1, Texas Instruments). The shaker input voltage~was adjusted via inverse filtering to maintain a near frequency-independent excitation, either in position or acceleration, depending on the experiment (see Figures~S12.C and~S12.D, Supporting Information).

\threesubsection{Acceleration Measurements and Processing}\\
Accelerations on opposing sides of the metamaterial were measured using two analog MEMS accelerometers (KXTC9-2050, Kionix) with a sensitivity of \SI[inter-unit-product = \!\cdot\!]{67.3}{\milli\volt\per\meter\second\squared}. Only the normal axis of each accelerometer was used, offering a mechanical bandwidth of \SI{1.8}{\kilo\hertz}. To prevent mechanical coupling, the accelerometers were connected through custom-made, highly flexible circuits cut from \SI{100}{\micro\meter}-thick polyimide films (Kapton, DuPont). The source-side accelerometer was affixed to the vibrating stage using cyanoacrylate adhesive (420, Loctite), while the output-side accelerometer was temporarily attached to the metamaterial with silicone adhesive (SI 5366, Loctite), cured at \SI{30}{\celsius} for \SI{30}{\minute}. Analog outputs were passed to an anti-aliasing stage implemented with low-noise, low distortion operational amplifiers (AD8652, Analog Devices) in a Sallen-Key topology (see Figure~S13, Supporting Information). This emulated a second-order Butterworth low-pass filter with a $\SI{1.5}{\kilo\hertz}$ cut-off. Signals were digitized at $\SI{20}{\kilo\hertz}$ by a 16-bit acquisition card (PCI-6221, National Instrument), detrended, and filtered by a zero-lag low-pass filter with a \SI{1}{\kilo\hertz} cut-off, achieving a noise floor of \SI[inter-unit-product = \!\cdot\!]{16}{\milli\meter\per\second\squared} RMS. Spectral responses were obtained via fast Fourier transforms (FFT), averaged across 10 trials, and smoothed using a Savitzky-Golay filter (50-sample window) to preserve sharp~peaks.

\threesubsection{Optical Vibrometry of the Unit Cells}\\
The standalone unit cells were backlit by a current-controlled white light source (CXB3590, Cree LED), resulting in highly contrasted images. To minimize specular reflections and improve image segmentation, a light-absorbent coating (AT205, Advance Tapes) was applied around each unit cell. Videos were recorded at \SI{7000}{\fps} with a monochrome high-speed CMOS camera (MotionBLITZ EoSens mini2, Mikrotron) at a resolution of $400 \times 400$ pixels. The camera was equipped with a \SI{50}{\milli\metre} lens (1-19558, Navitar) set to an $\mathrm{f}/2.8$ aperture. Although our unit cells are highly compliant, vibrations induced only minute changes in pixel intensity, invisible to the naked eye. To reveal these otherwise imperceptible motions, a phase-based amplification technique was applied~\cite{WadhwaEtAl-13}. A fiftyfold amplification was found empirically to provide optimal signal-to-noise ratio. For a harmonic excitation at a frequency, $f$, only variations within a narrow band, $[0.9f,1.1f]$, were amplified. The enhanced recordings were then binarized using Otsu's method, followed by morphological closing, i.e. dilation and erosion, using a disk-shaped structuring element with a diameter of $30$ pixels.

\threesubsection{Optical Vibrometry of the Metamaterial}\\
The optical setup was adjusted to image waves onto the metamaterial outer faces rather than within its individual cells. The frame rate was lowered to \SI{4000}{\fps}, allowing higher spatial resolution of $576\! \times\! 576$ pixels and detailed capture of the metamaterial. Bright-field illumination was achieved by positioning the white LED source in a near-coaxial, front-facing orientation. This setup highlighted surface features such as edges and defects, providing reliable tracking points. The processing workflow involved fiftyfold phase-based amplification, followed by flat-field correction (FFC) to compensate for nonuniform illumination (see Figure~S14, Supporting Information). The particle velocity field~$\boldsymbol{v}$ was estimated via optical flow with the Lucas-Kanade method. Assuming a continuously differentiable field, spatial filtering with a $40$-pixel Gaussian kernel was applied to mitigate discrete tracking artifacts. For the layer-by-layer analyses in Sections~\ref{sec:band_gap} and~\ref{sec:attenuation}, relevant regions were isolated using binary masks generated via morphological operations.

\newpage
\section*{Supporting Information}\vspace{-3mm}
Supporting Information is available from the Wiley Online Library or from the author.

\section*{Acknowledgments}\vspace{-3mm}
The authors thank Zhenishbek Zhakypov for his advice on the fabrication of 3D-printed channels. T.D. acknowledges funding from Fulbright National PhD Fellowship. This work was further supported by the ANR 20-CE33-0013 Maptics. 

\section*{Conflict of Interest}\vspace{-3mm}
The authors declare no conflict of interest.

\section*{Author Contributions}\vspace{-3mm}
T.D. conceived the research, designed and manufactured the metamaterial, conducted analytical, numerical, and experimental studies, analyzed the data, and wrote the initial draft of the manuscript. T.D., D.G., V.H., A.M.O., and S.H. secured fundings and supervised the work. T.D., D.G., A.M.O., and S.H. reviewed the manuscript.

\section*{Data Availability Statement}\vspace{-3mm}
The data that support the findings of this study are available from the corresponding author upon reasonable request.

\newpage
\bibliographystyle{MSP}
\bibliography{biblio_soft_metamaterial}

\end{document}


\pagestyle{fancy}
\rhead{\includegraphics[width=2.5cm]{vch-logo.png}}
\lhead{}

\title{Supporting Information}
\maketitle

\author{Thomas Daunizeau\textsuperscript{*}}
\author{David Gueorguiev}
\author{Vincent Hayward}
\author{Allison Okamura}
\author{Sinan Haliyo}
\dedication{}

\begin{affiliations}
T. Daunizeau, D. Gueorguiev, V. Hayward (deceased), S. Haliyo \par
Sorbonne Université, CNRS, ISIR, F-75005 Paris, France\par
E-mail: thomas.daunizeau@sorbonne-universite.fr\par
A.M. Okamura\par
Department of Mechanical Engineering, Stanford University, Stanford, CA 94305, USA\par
\end{affiliations}

\section{Theoretical Model}
\label{sec:theory}

\subsection{Mesofluidic Channels}
\label{sec:mesofluidic}

\begin{figure}[!t]
\centering
\includegraphics[width=178mm]{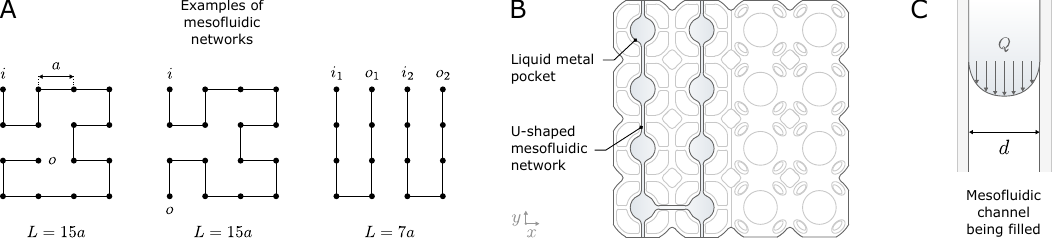}
\caption{A) Subset of paths visiting every node of a $4\!\times\!4$ grid. Left and middle Hamiltonian paths are impractically long,~with the left failing to terminate at the periphery. The right panel depicts a practical alternative using two shorter U-shaped paths. B) Implementation of these U-shaped~paths in our metamaterial. C) Cross-section of a mesofluidic channel under laminar flow.}
\label{fig:supporting_mesofluidic}
\end{figure}

The simplest approach for distributing liquid within the metamaterial is to use a unique channel that passes through each unit cell exactly once. This is, in fact, a classic graph theory problem~in which the nodes and edges correspond to the spherical pockets and channels, respectively. A solution to this problem is called a Hamiltonian path. It is trivial that a rectangular grid of nodes possesses Hamiltonian paths. Finding them all, however, is an NP-complete problem~\cite{CollinsKrompart-97}, which can be~extended to three dimensions. Two such paths are exemplified in \textbf{Figure~\ref{fig:supporting_mesofluidic}}.A for a $4\!\times\!4$ grid, corresponding to a cross-section of the acoustic metamaterial. The segments connecting the nodes must be orthogonal, that is, diagonal links are not allowed, since the mesofluidic channels can only be routed through the flexures. Two specific nodes are required: one input, $i$, for liquid injection and one output, $o$, for air removal. For practical reasons, they must be accessible from the periphery, which reduces the number of viable Hamiltonian paths. During the injection, the viscous friction against the walls of the mesofluidic channels causes an increase in upstream pressure. It is critical to keep it below a pressure threshold, $p_c$, to avoid catastrophic rupture of either the channels or the spherical shells. The Reynolds number, $\mathrm{Re}$, in a cylindrical channel is given by
\begin{equation}\label{eq:reynold_number}
\mathrm{Re} = \frac{4 \:\! \rho_l \:\! Q}{\mu_l \pi d}
\, ,
\end{equation}
where $\rho_l$ is the density of the liquid, $Q$ is the flow rate, $\mu_l$ is the dynamic viscosity, and $d$ is the internal diameter, as shown in Figure~\ref{fig:supporting_mesofluidic}.C. Assuming a manual injection of Galinstan ($\mu_l=\SI[inter-unit-product = \!\cdot\!]{2.4}{\milli\pascal\second}$) with a syringe at an estimated maximum flow rate of $\SI[inter-unit-product = \!\cdot\!]{10}{\micro\liter\per\second}$\!, through an internal diameter of $\SI{0.9}{\milli\meter}$, results in a Reynolds number $\mathrm{Re}\approx38$. Since $\mathrm{Re}<2000$, the flow is considered laminar~\cite{Reynolds-83}, and the pressure drop, $\Delta p$, along the mesofluidic channel is expected to follow the Hagen-Poiseuille law,
\begin{equation}\label{eq:hagen_poiseuille}
\Delta p = \frac{128\mu_l L Q}{\pi d^4}
\, ,
\end{equation}
where $L$ is the channel length. Therefore, the critical channel length, $L_c$, that must not be exceeded~is
\begin{equation}\label{eq:channel_critical_length}
L_c = \frac{\pi p_c d^4}{ 128\mu_l Q}
\, .
\end{equation}
Although the critical pressure, $p_c$, was not directly measured, trial and error indicated that for a lattice of size $a=\SI{16.5}{\milli\meter}$, a channel length of $L=15a=\SI{247.5}{\milli\meter}$ was too long whereas a length of $L=7a=\SI{115.5}{\milli\meter}$ stayed within safety limits. This sets a practical bound to the Hamiltonian paths shown in Figure~\ref{fig:supporting_mesofluidic}.A. The mesofluidic network was thus partitioned into eight shorter U-shaped channels, as illustrated in Figure~\ref{fig:supporting_mesofluidic}.B, enabling the supply of liquid metal to all $4^3$ unit cells.

\subsection{Unit Cell Lumped Dynamics}
\label{sec:lumped_model}

We developed a parsimonious model of the unit cell to estimate the dimensions required~for~a~low-frequency, subwavelength band gap. The geometry in \textbf{Figure~\ref{fig:supporting_lumped_model}}.A was simplified to~a~1D oscillator with resonance frequency $\omega_0=\sqrt{K/m}$, nested inside an outer shell of mass $M$.~As~illustrated in Figure~\ref{fig:supporting_lumped_model}.B, under the effective medium approximation, the unit cell is treated as~a~homogeneous mass $m^*\!$ which, to satisfy the dynamic equilibrium $F=\ddot{y}\,m^*$, is a function of $\omega$ such that, $\forall \, \omega \neq \omega_0$,
\begin{equation}\label{eq:effective_mass}
m^* = M+\frac{K}{{\omega_0}^{\!2}-{\omega}^{2}} 
\, ,
\end{equation}
with $\omega_0=2\pi f_0$. As shown in Figure~2.D and Figure~\ref{fig:supporting_lumped_model}.C, the effective mass $m^*\!$ is negative within a frequency band ranging from $f_0$ to $f_0\sqrt{1\!+\!m/M}$. Here, $m$ represents the mass of liquid metal contained in a spherical pocket of inner diameter $\phi$, given by
\begin{equation}\label{eq:fluid_mass}
m = \frac{\pi \rho_l r^3 a^3}{6}
\, ,
\end{equation}
where $\rho_l$ is the liquid metal density, $a$ is the lattice constant, and $r$ is the dimensionless ratio defined as $r=\phi/a$. The outer shell mass was estimated as $M=\rho_s V_{\!s}$, using the soft resin density $\rho_s$ and a computer-aided design (CAD) volume $V_{\!s}$ that captures its complex geometry. For a lattice of size $a=\SI{16.5}{\milli\meter}$, the shell volume is $V_{\!s}=\SI{918}{\milli\meter\cubed}$\!. Other sizes follow $V_{\!s}\propto a^3$\!, as indicated~in~Figure~\ref{fig:supporting_lumped_model}.E.

\subsection{Integration of Continuum Mechanics}

\begin{figure}[!t]
\centering
\includegraphics[width=178mm]{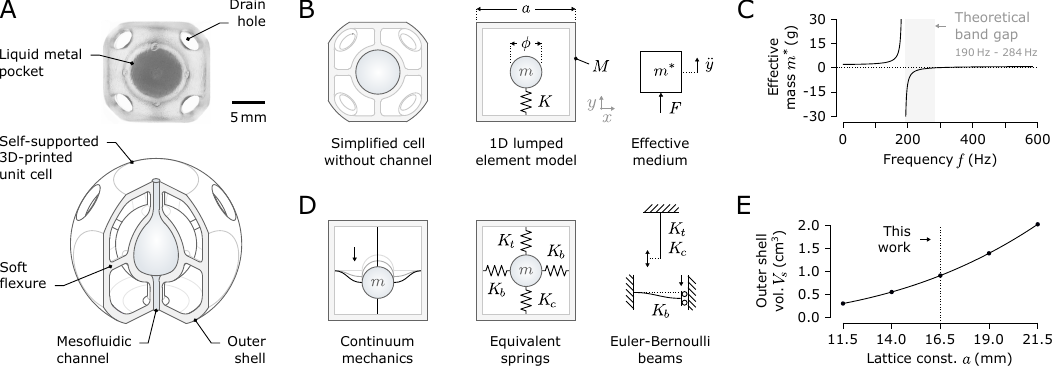}
\caption{A) Front-view photograph and CAD cross-section of the fully soft unit cell. B) Lumped 1D mass-spring model~and~the corresponding effective medium approximation as a homogeneous mass. C) Frequency-dependent effective mass for optimized dimensions $a=\SI{16.5}{\milli\meter}$ and $r=0.42$. D) Continuum mechanics model of the soft rods, treated as a parallel spring network of cantilevers under axial loading and fixed-guided beams under bending. E) Outer shell volume derived from the CAD geometry.}
\label{fig:supporting_lumped_model}
\end{figure}

The band gap limits are governed by the local resonance frequency $f_0$. To relate this frequency to the geometric parameters $a$ and $r$, we integrated continuum mechanics into the lumped-element framework. The internal flexure consists of six soft rods modeled as plain cylinders. The anisotropy caused by mesofluidic channels is omitted here and addressed separately through numerical simulations. Four of these rods lie in the plane orthogonal to motion, forming a linear guide. Following Euler-Bernoulli beam theory, they were modeled as fixed-guided beams, as illustrated in Figure~\ref{fig:supporting_lumped_model}.D. Accordingly, the bending stiffness $K_{\:\!\!b}$ of each of these four rods is
\begin{equation}\label{eq:bending_stiffness_soft}
K_{\:\!\!b} = \frac{3\pi ED^4}{2[a\,\xi(1-r)]^3}
\, ,
\end{equation}
where $E$ is Young's modulus, $D$ is the rod diameter, and $\xi$ is a geometric correction factor accounting for wall thickness and edge fillets.

\newpage
\noindent
The two remaining rods are aligned with the direction of motion and undergo alternating tension and compression. Provided that dynamic loads remain well below Euler’s critical load, buckling is precluded. Assuming so, the axial stiffnesses in tension, $K_{\:\!\!t}$, and compression, $K_{\:\!\!c}$, are~equal~and~given~by
\begin{equation}\label{eq:bending_stiffness_soft}
K_{\:\!\!t} = K_{\:\!\!c} = \frac{\pi ED^2}{2a\,\xi(1-r)}
\, .
\end{equation}
As depicted in Figure~\ref{fig:lumped_model}.D, treating these elements as a parallel spring network yields a total stiffness $K=4K_{\:\!\!b}+K_{\:\!\!t}+K_{\:\!\!c}$. The resonance frequency, $f_0$, of the lumped mass-spring system was thus derived from $\omega_0=\sqrt{K/m}$ as
\begin{equation}\label{eq:soft_local_resonance}
f_0 = \frac{1}{2\pi}\eta(r)\frac{D}{a^2}\sqrt{\frac{6E}{\xi\,\rho_l}}
\, ,
\end{equation}
where $\eta$ is defined by
\begin{equation}\label{eq:function_eta}
\eta(r) = \sqrt{\frac{1+6D^2/[a\,\xi(1-r)]^2}{r^3(1-r)}}
\, .
\end{equation}
The expression for $f_0$ in equation~\eqref{eq:soft_local_resonance} defines the design space in Figure~2.F, computed~with $D=\SI{1}{\milli\meter}$, $E=\SI{1.8}{\mega\pascal}$, and a geometric correction factor $\xi = 0.47$ extracted from the CAD model. Enhancing subwavelength performance requires an optimal ratio $r_{\!o}$ that minimizes $\eta$, defined in equation~\eqref{eq:function_eta}. Because this minimization has no closed-form solution, the optimal trajectory, shown as the dotted curve in Figure~2.F, was determined numerically through a golden-section search. Note that in the limit where the ratio $(D/a)^2$ is negligible, the function simplifies to
\begin{equation}\label{eq:limit_phi}
\lim_{(D/a)^2 \to \, 0}\eta(r) = \sqrt{\frac{1}{r^3(1-r)}}
\, ,
\end{equation}
which is minimal for $r_{\!o}=3/4$. In this study, however, manufacturing limits of soft SLA restricted the ratio $r$ to a practical range of $[0.30,0.55]$.

\newpage
\section{Finite Element Model}
\label{sec:finite_elements}

While the lumped model captured a single degree of freedom, additional vibration modes arise from geometric features previously omitted, such as mesofluidic channels, rounded edges, and drain holes. Finite strains, solid-liquid coupling, and nonlinear material properties further call for finite element analysis to tune the metamaterial. A refined 3D model grounded in continuum mechanics and fluid dynamics was thus derived. It is divided into three parts: a hyperelastic lattice, a liquid infill, and boundary conditions. In the following, the subscripts \textit{s} and \textit{l} stand for solid and liquid, respectively.

\subsection{Hyperelastic Lattice}
\label{sec:hyperelastic}

The soft resin lattice was assumed to be homogeneous and isotropic. When excited at resonance, its high flexibility establishes large strains that could not be captured by linear elasticity. Instead, it was modeled as a nearly incompressible Neo-Hookean hyperelastic material, defined by the strain energy density function, $W \! \in \mathbb{R}$, such as
\begin{equation}\label{eq:neo_hookean}
W = \frac{\mu}{2}(I_1J^{-2/3}-3)+\frac{\kappa}{2}(J-1)^2 
\, ,
\end{equation}
with $I_1=\tr(\mathbf{C}) \in \mathbb{R}$ the first invariant of the right Cauchy-Green strain tensor, $J=\det(\mathbf{F}) \in \mathbb{R}$ the third invariant of the deformation gradient tensor, $\mu \in \mathbb{R}_{>0}$ Lamé's second parameter, and $\kappa \in \mathbb{R}_{>0}$ the bulk modulus, given by
\vspace*{-2mm}
\begin{equation}\label{eq:lame_bulk}
\mu = \frac{E}{2(1+\nu)} \quad \textrm{and} \quad \kappa = \frac{E}{3(1-2\nu)} 
\, ,
\end{equation}
where $E \in \mathbb{R}_{>0}$ is Young's modulus and $\nu \in [0,0.5[$ is Poisson's ratio. The right Cauchy-Green strain tensor, $\mathbf{C} \in \mathbb{R}^{3 \times 3}$, the Green-Lagrange strain tensor, $\mathbf{E} \in \mathbb{R}^{3 \times 3}$, and the deformation gradient tensor, $\mathbf{F} \in \mathbb{R}^{3 \times 3}$, are all function of the displacement field, $\boldsymbol{u_s} \in \mathbb{R}^3$, such as
\begin{gather}
\mathbf{C}=2\mathbf{E}+\mathbf{I}
\, ,
\\
\mathbf{E}=\frac{1}{2}\big[(\nabla\boldsymbol{u_s})^{\mathrm{T}}+\nabla\boldsymbol{u_s}+(\nabla\boldsymbol{u_s})^{\mathrm{T}} \cdot \nabla\boldsymbol{u_s}\big]
\, ,
\\
\mathbf{F}=\nabla\boldsymbol{u_s}+\mathbf{I}
\, ,
\end{gather}
where $\mathbf{I}$ is the identity matrix. Under the finite strain theory, the dynamic equilibrium yields the following wave equation,
\begin{equation}\label{eq:wave_large_deformation}
\rho_s \, \pdv[2]{\boldsymbol{u_s}}{t} = \nabla\bigg(\!\mathbf{F} \pdv[]{W}{\boldsymbol{E}}\bigg)^{\!\!\mathrm{T}}
\, ,
\vspace*{2mm}
\end{equation}
where $\rho_s$ is the density. The separation of variables provides a general solution decomposed in spatial and temporal components as in $\boldsymbol{u_s}(\boldsymbol{x},t) = \boldsymbol{\tilde{u}_s}(\boldsymbol{x}) e^{i \omega t}$, where $\boldsymbol{x} = [x \, y \, z] \in \mathbb{R}^3$ is the coordinate vector, $\boldsymbol{\tilde{u}_s} \in \mathbb{C}$ is the complex amplitude, and $\omega$ is the angular frequency. Equation~\eqref{eq:wave_large_deformation} can thus be solved in the frequency domain as
\begin{equation}\label{eq:wave_solid_harmonic}
- \rho_s \, \omega^2 \boldsymbol{\tilde{u}_s} = \nabla\bigg(\!\mathbf{F} \pdv[]{W}{\boldsymbol{E}}\bigg)^{\!\!\mathrm{T}}
\, .
\end{equation}

\newpage
\subsection{Fluid Dynamics}
\label{sec:fluid}

Galinstan was assumed to be homogeneous, isotropic, irrotational and Newtonian, i.e. with~a~viscosity independent of strain rate. As net flow between unit cells ceased upon the sealing of the mesofluidic channels, the fluidic domain was reduced to that of a single unit cell. Conservation of mass in an elementary volume of fluid is given by the continuity equation,
\begin{equation}\label{eq:continuity_liquid}
\pdv[2]{p}{t} + c_l^2 \rho_l \nabla \pdv[]{\boldsymbol{v_l}}{t}= 0
\, ,
\end{equation}
where $p \in \mathbb{R}$ is the pressure, $\rho_l$ is the average density of the liquid, $c_l$ is the speed of sound, and $\boldsymbol{v_l} \in \mathbb{R}^3$ is the particle velocity. Fluid motion is governed by the linearized Navier-Stokes equation,
\begin{equation}\label{eq:linear_navier_stokes}
-\pdv[]{\boldsymbol{v_l}}{t}= \frac{1}{\rho_l}\nabla p + \Big(\frac{4}{3}\mu+\lambda \Big) \frac {1}{(c_l\:\!\rho_l)^2} \nabla \pdv[]{p}{t}
\, ,
\end{equation}
where $\mu_l$ is the dynamic viscosity and $\lambda$ is the bulk viscosity. Combining both equations~\eqref{eq:continuity_liquid} and~\eqref{eq:linear_navier_stokes} yields the following wave equation,
\begin{equation}\label{eq:wave_liquid}
\frac{1}{c_l^2 \rho_l} \, \pdv[2]{p}{t} = \frac{1}{\rho_l}\nabla^2 p + \Big(\frac{4}{3}\mu_l+\lambda \Big) \frac {1}{(c_l\:\!\rho_l)^2} \nabla^2 \pdv[]{p}{t}
\, .
\end{equation}
The liquid metal can further be assumed incompressible, so that isotropic dilatations of an elementary volume of liquid cannot induce viscous stresses, hence $\lambda=0$. Separation of variables, similar to that previously applied to the soft lattice, gives $p=\tilde{p}({x},t) e^{i \omega t}$ with $\tilde{p} \in \mathbb{C}$ the complex amplitude. In turn, equation~\eqref{eq:wave_liquid} becomes an inhomogeneous Helmholtz equation,
\begin{equation}\label{eq:inhomogeneous_helmholtz}
-\frac{\omega^2}{c_l^2\rho_l} \tilde{p} = \frac{1}{\rho_l}\nabla^2 \tilde{p} + \frac{4}{3}\mu_l \frac{i \omega}{(c_l\:\!\rho_l)^2} \nabla^2 \tilde{p}
\, .
\end{equation}
Because the fluid domain is formulated in terms of particle velocity, the deformed fluid geometry shown in grayscale in Figure~3.A is not available directly. Therefore, an effective displacement field of the fluid, $\boldsymbol{u_l} \in \mathbb{R}^3$, was estimated from the particle velocity using the harmonic kinematic relation, $\boldsymbol{u_l}={\boldsymbol{v_l}}/{i \omega}$. When the displacement field is scaled up for visualization, this reconstruction can lead to a small apparent mismatch at the solid-liquid interface.

\subsection{Boundary Conditions}
\label{sec:boundaries}

The solid-liquid interface is defined by the smooth boundary $\partial\Omega_{\,\mathrm{SL}}$ described in Figure~2.B. Continuity of the displacement field under a no-slip condition entails that $\left. \boldsymbol{\tilde{u}_l} \right|_{\partial\Omega_{\,\mathrm{SL}}} = \left. \boldsymbol{\tilde{u}_s} \right|_{\partial\Omega_{\,\mathrm{SL}}}$. Local dynamic equilibrium must also be satisfied, hence
\begin{equation}\label{eq:local_coupling}
- \rho_s \, \omega^2 \left. \boldsymbol{\tilde{u}_s} \right|_{\partial\Omega_{\,\mathrm{SL}}}\! \cdot \boldsymbol{n} = \left. \nabla \tilde{p} \, \right|_{\partial\Omega_{\,\mathrm{SL}}}
\, ,
\end{equation}
where $\boldsymbol{n} \in \mathbb{R}^3$ is the unit vector normal to the boundary $\partial\Omega_{\,\mathrm{SL}}$. To reveal the band structure, the cubic lattice was considered to be of infinite 3D periodicity. The problem could thus be reduced to a single unit cell with periodic boundaries applied to each side of the cubic lattice. The remaining surfaces were free. In these terms, the Floquet-Bloch's theorem implies that solutions of the wave equations in both solid and fluidic domains are in the form
\begin{equation}\label{eq:bloch_theorem_solid}
\boldsymbol{\tilde{u}_s}(\boldsymbol{x} + a\boldsymbol{q}) = \boldsymbol{\tilde{u}_s}(\boldsymbol{x}) \, e^{i a \boldsymbol{k}\cdot \boldsymbol{q}},
\end{equation}
\begin{equation}\label{eq:bloch_theorem_liquid}
\boldsymbol{\tilde{u}_l}(\boldsymbol{x} + a\boldsymbol{q}) = \boldsymbol{\tilde{u}_l}(\boldsymbol{x}) \, e^{i a \boldsymbol{k}\cdot \boldsymbol{q}},\vspace*{2mm}
\end{equation}
where $\boldsymbol{x} = [x \, y \, z]^\mathrm{T}\! \in \mathbb{R}^3$ is the position vector, $\boldsymbol{q} \in \mathbb{Z}^3$ is the iteration vector, $\boldsymbol{k} = [k_x \, k_y \, k_z]^\mathrm{T}\! \in \mathbb{R}^3$ is the wave vector. In the reciprocal space ($\boldsymbol{k}$-space) of a cubic lattice, the irreducible Brillouin zone (IBZ) is a tetrahedron bounded by the critical points $\{\Gamma, \mathrm{X}, \mathrm{M}, \mathrm{R}\}$, as schematized in Figure~2.E. Dispersion diagrams were obtained by sweeping the wave vector $\boldsymbol{k}$ along the periphery of the IBZ, as detailed in \textbf{Table~\ref{tab:sweep_sequence}}. To maintain continuity, the edge $\mathrm{X}\mathrm{M}$ had to be scanned twice. This is because the 1-skeleton of a tetrahedron does not possess a Eulerian path.

\begin{table}[!h]
    \centering
    \small
    \renewcommand{\arraystretch}{1.2} 
    \setlength{\tabcolsep}{4mm} 
    \caption{FEA sweep sequence for computing band structure across the IBZ.}
    \label{tab:sweep_sequence}
	\begin{threeparttable}
    	\begin{tabular}{ccrrr}
        \toprule
    	Edge & Condition & \textbf{$k_x$} & \textbf{$k_y$} & \textbf{$k_z$}\\	
    	\midrule[0.4pt]
    		$\Gamma\mathrm{X}$     & $0 \leq p  <  1$ & 0            & $\pi p/a$    & 0            \\
    		$\mathrm{X}\mathrm{M}$ & $1 \leq p  <  2$ & $\pi(p-1)/a$ & $\pi/a$      & 0            \\
    		$\mathrm{M}\Gamma$     & $2 \leq p  <  3$ & $\pi(3-p)/a$ & $\pi(3-p)/a$ & 0            \\
    		$\Gamma\mathrm{R}$     & $3 \leq p  <  4$ & $\pi(p-3)/a$ & $\pi(p-3)/a$ & $\pi(p-3)/a$ \\
    		$\mathrm{R}\mathrm{X}$ & $4 \leq p  <  5$ & $\pi(5-p)/a$ & $\pi/a$      & $\pi(5-p)/a$ \\
    		$\mathrm{X}\mathrm{M}$ & $5 \leq p  <  6$ & $\pi(p-5)/a$ & $\pi/a$      & 0            \\
    		$\mathrm{M}\mathrm{R}$ & $6 \leq p\leq 7$ & $\pi/a$      & $\pi/a$      & $\pi(p-6)/a$ \\
        \bottomrule
        \end{tabular}
    \end{threeparttable}
\end{table}

\begin{figure}[!b]
\centering
\includegraphics[width=178mm]{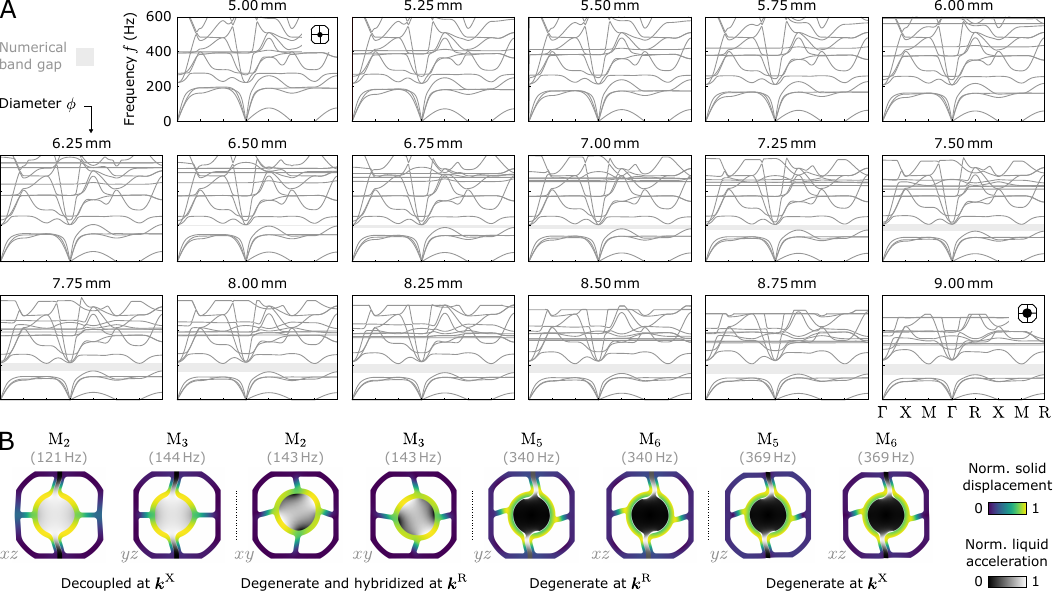}
\caption{A) Complete dispersion diagram dataset from finite element analyses for liquid metal inclusion diameters $\phi$ ranging from \SI{5}{\milli\meter} to \SI{9}{\milli\meter}, corresponding to ratios $r$ of $0.30$ to $0.55$. Only branches for the first sixteen vibration modes are illustrated. B) Additional local resonances of interest involving degeneration and/or hybridization, presented for wave vectors $\boldsymbol{k}^\mathrm{R}$\! and $\boldsymbol{k}^\mathrm{X}$\!.}
\label{fig:supporting_dataset_fea}
\end{figure}

\subsection{Extended Analysis of Dispersion Relations and Local Resonances}
\label{sec:dispersion}

To optimize the metamaterial geometry, we performed iterative numerical simulations by varying the diameter $\phi$ of the liquid metal pocket. The results reported in Figure~2.C were obtained~by~aggregating the dispersion graphs in \textbf{Figure~\ref{fig:supporting_dataset_fea}}.A. After fine-tuning the unit cell, we analyzed the local resonances depicted in Figure~3.A to identify the vibration modes driving band gap formation.

\newpage
\noindent
Additional modes of interest are illustrated in Figure~\ref{fig:supporting_dataset_fea}.B. Specifically, as the wave vector transitions from $\boldsymbol{k}^\mathrm{X}$\! to $\boldsymbol{k}^\mathrm{R}$, modes $\mathrm{M}_2$ and $\mathrm{M}_3$ become degenerate at \SI{143}{\hertz}, forming a hybrid mode characterized by diagonal motion of the liquid metal. In contrast, from $\boldsymbol{k}^\mathrm{R}$\! to $\boldsymbol{k}^\mathrm{X}$, modes $\mathrm{M}_5$ and $\mathrm{M}_6$ reach~a~distinct degenerate state at \SI{369}{\hertz} without hybridization.

\subsection{Comparison of Solid and Liquid Inclusions}
\label{sec:solid_vs_liquid}

Results in Figure~3.A show that torsion mobilizes only a negligible fraction of the liquid metal inertia. We hypothesize that this weak inertial contribution is key for opening the band gap. To test this, we performed numerical simulations replacing the liquid inclusion with a solid inclusion of equal density of $\SI[inter-unit-product = \!\cdot\!]{6.44}{\gram\per\centi\meter\cubed}$.\! The solid was modeled as a rigid linear elastic body with a Poisson's ratio~of~$0.3$ and a Young's modulus of $\SI{210}{\giga\pascal}$. To avoid geometry-induced artifacts, the two protruding stems of the liquid-filled channel in Figure~2.B were removed, leaving only a spherical inclusion. 

\begin{figure}[!h]
\centering
\includegraphics[width=178mm]{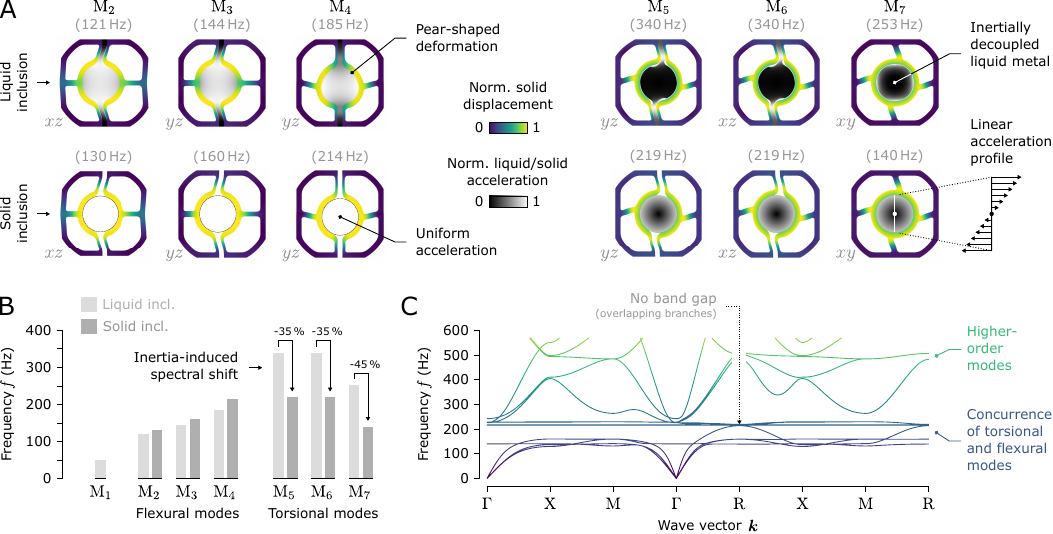}
\caption{A) Vibration modes for liquid (top) and solid (bottom) inclusions. During torsion, the liquid decouples from~the~lattice, leaving only a thin boundary layer contributing to inertia. In contrast, the solid core undergoes uniform angular acceleration, resulting in greater inertia. B) Resonance frequencies from FEA. The liquid inclusion yields significant spectral separation between flexion and torsion. C) Dispersion graph for a solid inclusion, characterized by overlapping branches and no band gap.}

\label{fig:supporting_solid_liquid_comparison}
\end{figure}

Modal analysis reveals that the breathing mode, $\mathrm{M}_1$, is suppressed when the inclusion is solid. As illustrated in \textbf{Figure~\ref{fig:supporting_solid_liquid_comparison}}.A, the flexural modes, $\mathrm{M}_2, \mathrm{M}_3, \mathrm{and}\, \mathrm{M}_4$, remain largely unaffected, with the solid inclusion undergoing uniform acceleration. The response differs markedly for the torsional modes $\mathrm{M}_5, \mathrm{M}_6, \mathrm{and}\, \mathrm{M}_7$. For a liquid inclusion, only a thin boundary layer follows the torsional~motion. Conversely, for a solid inclusion, the tangential acceleration scales linearly with radial distance from the rotation axis. In turn, the entire mass of the inclusion contributes to the moment of~inertia. Because the moment of inertia is significantly larger for the solid inclusion, the resonance frequencies of torsional modes shift downward while flexural modes remain nearly stationary, as shown in Figure~\ref{fig:supporting_solid_liquid_comparison}.B. For example, at wave vector $\boldsymbol{k}^\mathrm{R}$, mode $\mathrm{M}_7$ occurs at \SI{140}{\hertz} for the solid inclusion, 45\% lower than the \SI{253}{\hertz} observed in the liquid case. This unilateral shift eliminates the spectral separation between torsional and flexural branches. In turn, the band gap collapses, as evidenced by the overlapping branches shown in Figure~\ref{fig:supporting_solid_liquid_comparison}.C.

\section{Experimental Results}
\label{sec:results}

\subsection{State-Space Model of a Standalone Unit Cell}
\begin{figure}[!h]
\centering
\includegraphics[width=178mm]{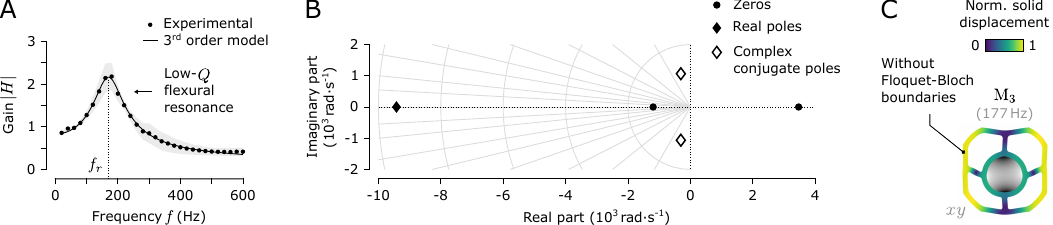}
\caption{A) Magnitude of the transfer function between liquid metal and outer shell displacement. Data are averaged over ten standalone unit cells and fitted to a third-order state-space model. The shaded area indicates $\pm 1$ SD. B) Pole-zero map of the identified system. C) Undamped flexural mode of a standalone unit cell, simulated without periodic boundary conditions.}
\label{fig:supporting_state_space_model}
\end{figure}

To bridge numerical predictions with empirical observations, the local resonance dynamics of~stand-alone unit cells were characterized. We extracted the output~$y_\mathrm{out}(t)$ and input $y_\mathrm{in}(t)$, defined as the liquid metal and outer shell centroids, respectively. We then derived the transfer function $H(s) = Y_\mathrm{out}(s)/Y_\mathrm{in}(s)$, with $s$ the Laplace variable. Experimental data were fitted to a third-order state-space model ($\mathrm{RMSE}=0.03$) using Levenberg-Marquardt optimization, as shown by the Bode plot reproduced in \textbf{Figure~\ref{fig:supporting_state_space_model}}.A. The state vector is defined as $\boldsymbol{\xi} = [\xi_1 \, \xi_2 \, \xi_3]^\mathrm{T}\! \in \mathbb{R}^3$, and the system dynamics are governed by the state-space equations,
\begin{align}\label{eq:state_space}
\dot{\boldsymbol{\xi}}(t) &= \mathbf{A}\,\boldsymbol{\xi}(t) + \mathbf{B}\,y_\mathrm{in}(t) \, , \\
y_\mathrm{out}(t) &= \mathbf{C}\,\boldsymbol{\xi}(t) + D\,y_\mathrm{in}(t) \, ,
\end{align}
with 
\begin{equation}\label{eq:matrices_ABCD}
\mathbf{A} = 
\begin{bmatrix} 
-305.5 & 1065 & 0 \\
-1065 & -305.5 & -2073 \\
0.01563 & -0.1285 & -9422 \\
\end{bmatrix}
\, , \,
\mathbf{B} = 
\begin{bmatrix} 
0 \\
18.26 \\
113.5 \\
\end{bmatrix}
\, , \,
\mathbf{C}^\mathrm{T} = 
\begin{bmatrix} 
-106.1 \\
-127.1 \\
-0.002 \\ 
\end{bmatrix}
\, , \,
D = 0
\, .
\end{equation}
This third-order system is governed by a real pole $p_1 = \SI[inter-unit-product = \!\cdot\!]{-9422}{\radian\per\second}$\! and a complex conjugate pair $p_{2,3} = -\alpha\,\pm \, i\,\omega_d = -305.5 \pm i\,1065\,\SI[inter-unit-product = \cdot]{}{\radian\per\second}$\!. The natural frequency $\omega_n = \sqrt{\alpha^2 + \omega_d^2} =\SI[inter-unit-product = \cdot]{1108}{\radian\per\second}$\! and damping ratio $\zeta = \alpha/\omega_n = 0.276$ were derived from the decay rate~$\alpha$ and damped frequency~$\omega_d$ of this oscillatory mode. The quality factor was derived as $Q = 1/2\,\zeta =1.81$. Two zeros were~identified at $z_1 = \SI[inter-unit-product = \!\cdot\!]{-1195}{\radian\per\second}$\! and $z_2 = \SI[inter-unit-product = \!\cdot\!]{3463}{\radian\per\second}$\!, as indicated by the pole-zero map in Figure~\ref{fig:supporting_state_space_model}.B. The corresponding transfer function is given by
\begin{equation}\label{eq:transfer_function}
H(s) = \frac{G\, (s - z_1)(s - z_2)}{(s - p_1)(s^2 + 2\,\zeta\omega_n s + \omega_n^2)} \, ,
\end{equation}
with the scaling factor $G=\mathbf{C}\mathbf{B}$. The static gain, $H(0)$, represents the steady-state ratio of liquid metal displacement to outer shell displacement. It is defined as $H(0) = D - \mathbf{C}\mathbf{A}^{-1}\mathbf{B} = -{G z_1 z_2}/{p_1 \omega_n^2}$. Substituting the experimental data yields $H(0) = 0.83$. While theory predicts a static gain of unity, this discrepancy is likely due to the limited fidelity of camera-based vibrometry at low frequencies.

Additional FEA was conducted without Floquet-Bloch boundary conditions to replicate the empirical conditions of a standalone unit cell. As shown in Figure~\ref{fig:supporting_state_space_model}.C, the FEA predicts~the~flexural mode $\mathrm{M}_3$ at \SI{177}{\hertz}. This closely matches the undamped natural frequency, $f_n = \omega_n / 2\pi = \SI{176}{\hertz}$, a relevant baseline since our simulations neglect dissipative effects. Such accuracy further validates the material properties and constitutive laws implemented in our finite element model.

\newpage
\subsection{Phase Velocity Estimation}
\label{sec:phase_velocity}

\begin{figure}[!h]
\centering
\includegraphics[width=178mm]{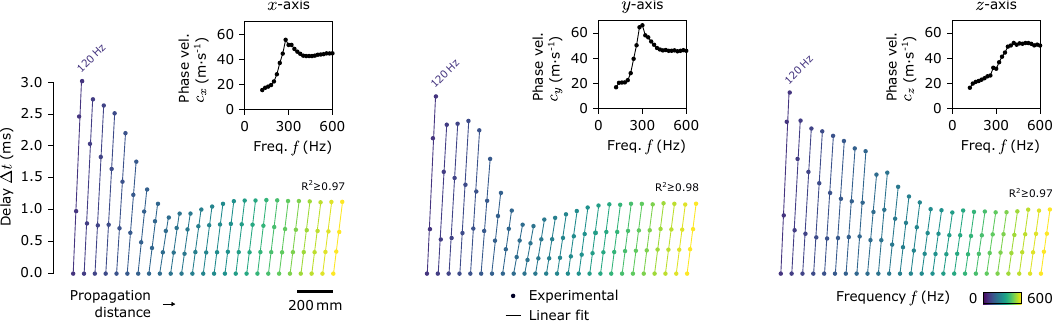}
\caption{Linear regression of wavefront time delays $\Delta t$ against propagation distance through successive unit cell layers. Each fit shows a distinct excitation frequency. Datasets are horizontally offset by \SI{60}{\milli\meter} for clarity. Values below \SI{120}{\hertz} are omitted~due to finite-size effects from excessively large wavelengths. Insets display the resulting frequency-dependent~phase~velocity.}
\label{fig:supporting_phase_velocity}
\end{figure}

Obtaining the empirical dispersion relations requires a precise estimate of the phase velocity. This was achieved by first measuring the particle velocity on the outer faces of the metamaterial via full-field optical vibrometry. Spatially averaging this field across successive layers of unit cells revealed the phase-lagged oscillations shown in Figure~4.B. The phase velocity, $\boldsymbol{c}$, was then derived through linear regression of the time delays, $\Delta t$, against propagation distance, as shown in Figure~4.C. The complete dataset, presented in \textbf{Figure~\ref{fig:supporting_phase_velocity}}, exhibits excellent agreement with the linear fits. The resulting frequency-dependent phase velocities for each principal axis are reported in the corresponding insets.

\vspace*{-4mm}
\subsection{Summary of Band Gap Properties}
\label{sec:band_gap}

We conducted a multifaceted investigation of the band gap using lumped-element modeling, finite element analysis, full-field optical vibrometry, and accelerometer measurements. The band gap~limits are summarized in \textbf{Table~\ref{tab:band_gap_results}}, where the complete band gap is $\mathrm{BG}_x \cap \mathrm{BG}_y \cap \mathrm{BG}_z$. Based on~optical data, peak attenuation along the principal axes $\boldsymbol{e}_i$ occurred at $\{240, 240, 320\}\,\SI{}{\hertz}$. At these frequencies, the real part of the dispersion relations yield wavelengths $\lambda_i=\{155, 164, 115\}\,\SI{}{\milli\meter}$, resulting in dimensionless ratios $\lambda_i/a=\{9.4,9.9,7.0\}$ that indicate operation in a deep subwavelength regime.

\begin{table}[!h]
    \centering
    \small
    \renewcommand{\arraystretch}{1.2} 
    \setlength{\tabcolsep}{4mm} 
    \caption{Summary of band gap frequency limits obtained from theoretical, numerical, and empirical methods.}
    \label{tab:band_gap_results}
    \begin{tabular}{
        l 
        S[table-format=3]
        S[table-format=3]
        S[table-format=3]
        S[table-format=3]
        S[table-format=3]
        S[table-format=3]
        S[table-format=3]
        S[table-format=3]
    }
        \toprule
        & \multicolumn{2}{c}{$\mathrm{BG}_x$\,(\SI{}{\hertz})} & \multicolumn{2}{c}{$\mathrm{BG}_y$\,(\SI{}{\hertz})} & \multicolumn{2}{c}{$\mathrm{BG}_z$\,(\SI{}{\hertz})} & \multicolumn{2}{c}{Complete (\SI{}{\hertz})} \\
        \cmidrule(lr){2-3} \cmidrule(lr){4-5} \cmidrule(lr){6-7} \cmidrule(lr){8-9}
        Method & {Lower} & {Upper} & {Lower} & {Upper} & {Lower} & {Upper} & {Lower} & {Upper} \\
        \midrule
        Theoretical (lumped)  & {--} & {--} & {--} & {--} & {--} & {--} & 190 & 284 \\
        Numerical (FEA)       & {--} & {--} & {--} & {--} & {--} & {--} & 185 & 208 \\
        Optical (dispersion)  & 180 & 280 & 200 & 280 & 260 & 380 & 260 & 280 \\
        Optical (attenuation) & 160 & 340 & 160 & 380 & 200 & 540 & 200 & 340 \\
        Accelerometer         & 120 & 330 & 130 & 340 & 210 & 600 & 210 & 330 \\
        \bottomrule
    \end{tabular}
\end{table}

\vspace*{-2mm}
\subsection{Extended Dataset of RMS Particle Velocity}
\label{sec:rms_dataset}

\begin{figure}[!h]
\centering
\includegraphics[width=178mm]{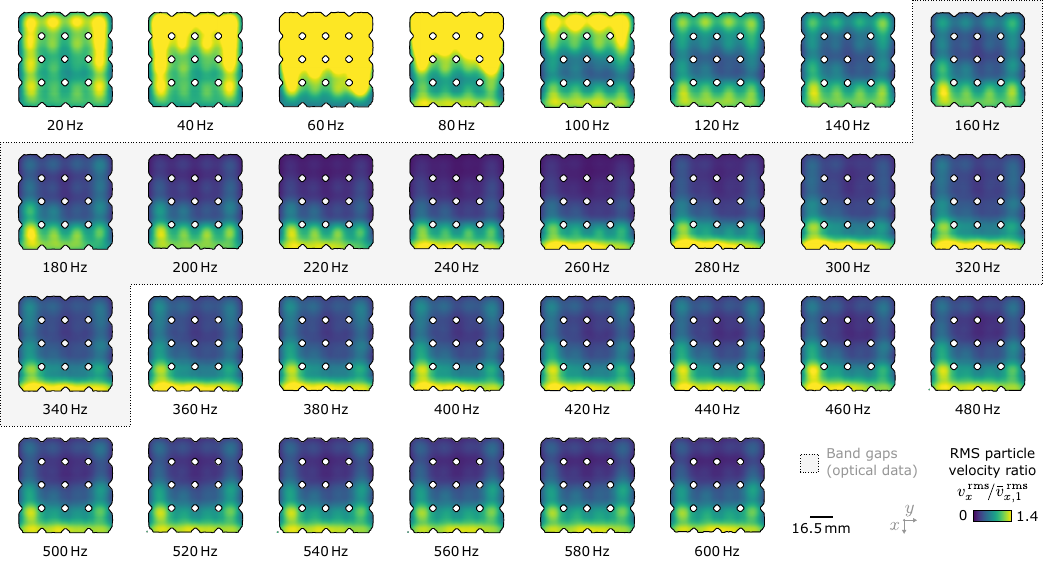}
\caption{RMS particle velocity field along the $x$-axis from \SI{20}{\hertz} to \SI{600}{\hertz}, normalized to the first layer average. Ratios below unity denote attenuation, while those above unity are clipped at 1.4 as bulk resonance falls beyond the scope of this study.}
\label{fig:supporting_RMS_particle_velocity_x}
\end{figure}

\begin{figure}[!h]
\centering
\includegraphics[width=178mm]{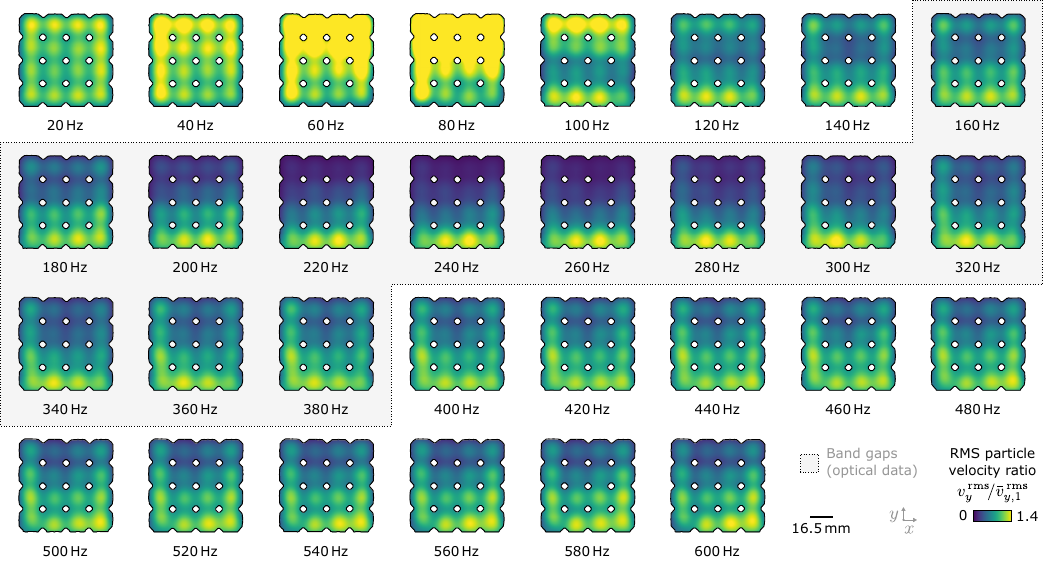}
\caption{RMS particle velocity field along the $y$-axis from \SI{20}{\hertz} to \SI{600}{\hertz}, normalized to the first layer average. Ratios below unity denote attenuation, while those above unity are clipped at 1.4 as bulk resonance falls beyond the scope of this study.}
\label{fig:supporting_RMS_particle_velocity_y}
\end{figure}

\begin{figure}[!h]
\centering
\includegraphics[width=178mm]{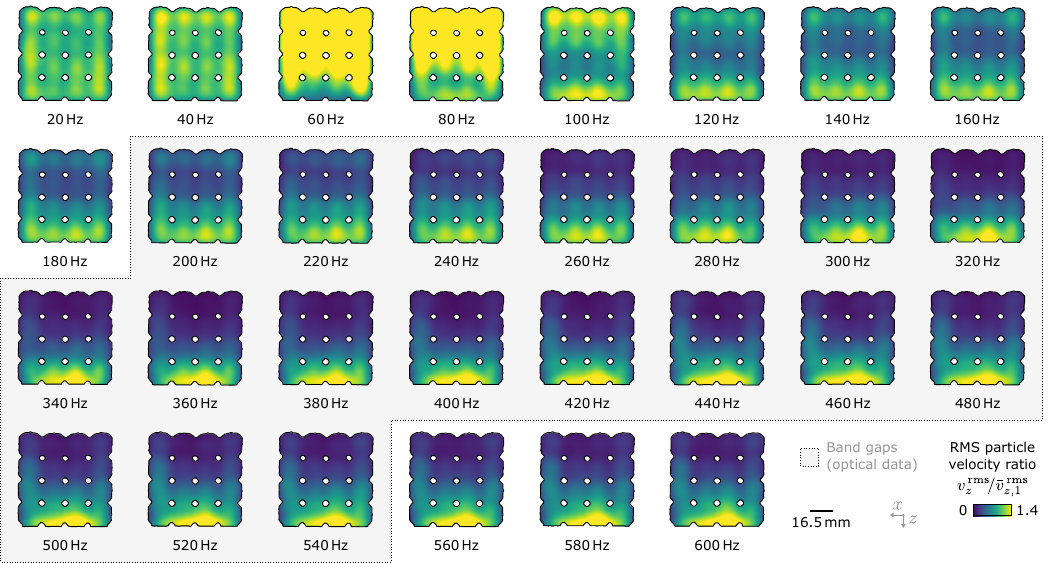}
\caption{RMS particle velocity field along the $z$-axis from \SI{20}{\hertz} to \SI{600}{\hertz}, normalized to the first layer average. Ratios below unity denote attenuation, while those above unity are clipped at 1.4 as bulk resonance falls beyond the scope of this study.}
\label{fig:supporting_RMS_particle_velocity_z}
\end{figure}

Motion amplification introduces a Eulerian-Lagrangian mismatch, as physical material points drift across fixed pixel coordinates during temporal integration. Consequently, the resulting RMS particle velocity field $\boldsymbol{v}^\mathrm{rms}$ is an approximation projected onto a static reference configuration, specifically the final frame. Its visualization in Figure~5.A was restricted to the $y$-axis and a set of representative frequencies. The full dataset is provided in \textbf{Figure~\ref{fig:supporting_RMS_particle_velocity_x}} for the $x$-axis, \textbf{Figure~\ref{fig:supporting_RMS_particle_velocity_y}} for the $y$-axis, and \textbf{Figure~\ref{fig:supporting_RMS_particle_velocity_z}} for the $z$-axis. Finally, these RMS fields were normalized by the spatial average of the first layer $\mathrm{L}_1$.\! This compensates for residual frequency-dependent variations in the input vibration that persisted despite the amplitude correction described in Section~S\ref{sec:shaker}.

\vspace*{-2mm}

\vspace*{-2mm}

\section{Experimental Setup and Prototyping}
\label{sec:setup}

\subsection{Stereolithography of a Soft Lattice}
\label{sec:stereolithography}

\begin{figure}[!t]
\centering
\includegraphics[width=178mm]{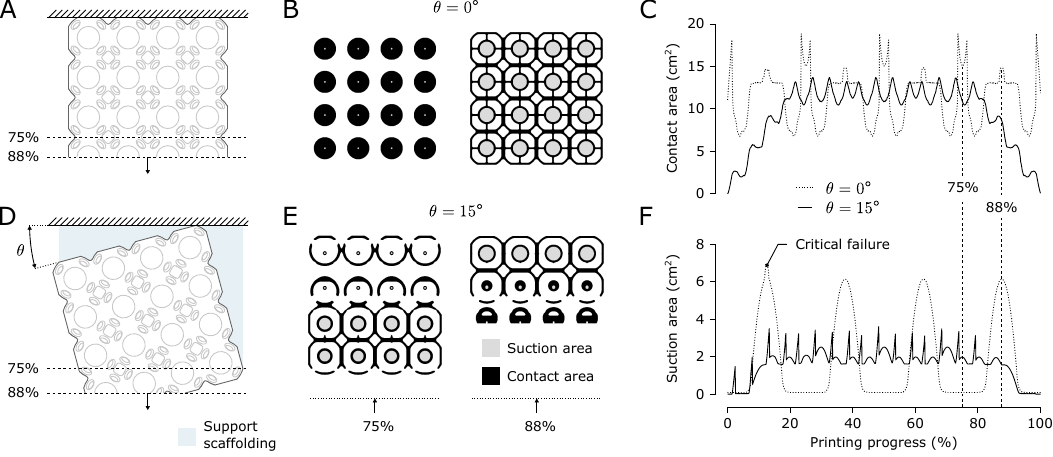}
\caption{A) Metamaterial lattice on the SLA build plate in an orthogonal orientation. B) Contact patterns across distinct printing stages for the orthogonal orientation. C) Evolution of contact area between the cured resin and the release film during printing. D) Optimized \SI{15}{\degree} tilt to minimize interfacial adhesion. E) Contact patterns for a \SI{15}{\degree} tilt. F) Evolution of the suction area, showing how tilting suppresses peaks and, in turn, prevents structural failures such as pocket puncturing.}
\label{fig:supporting_soft_stereolithography}
\end{figure}

In SLA 3D printing, each layer is formed in a cycle in which the part contacts a release film and UV light is focused at the film-resin interface to cure the resin. However, the spherical pocket in each unit cell acted as a suction cup, producing excessive adhesion forces when the build platform pulled away from the release film. As depicted in \textbf{Figures~\ref{fig:supporting_soft_stereolithography}}.A and~\ref{fig:supporting_soft_stereolithography}.B, printing along orthogonal axes can engage up to sixteen cups simultaneously. The resulting peaks in suction area, as shown in Figure~\ref{fig:supporting_soft_stereolithography}.F, caused frequent failures with punctured spherical pockets. Printing at an angle, $\theta$, mitigated this effect. Finding the best build orientation for a printed volume, $V$, amounts to minimizing the instantaneous suction area, $A_{\theta}$, defined by
\vspace*{-1mm}
\begin{equation}\label{eq:printing}
V = \int_{0}^{T}\!\! A_{\theta}(t)\, dt =  \mean{A}_{\theta}\, T 
\, ,
\end{equation}
where $T$ is the printing duration and $\mean{A}_{\theta}$ is the time-averaged suction area. By definition, $\max{(A_{\theta})}$ is minimal if $A_{\theta}\rightarrow\mean{A}_{\theta},\,\forall t\in[0,T]$. An angle $\theta=\SI{15}{\degree}$, as depicted in Figure~\ref{fig:supporting_soft_stereolithography}.D, was found~empirically to be optimal. As shown in Figures~\ref{fig:supporting_soft_stereolithography}.E and~\ref{fig:supporting_soft_stereolithography}.F, it reduced peaks in suction area by roughly~44\%. A similar strategy can be used to reduce the adhesive contact area between freshly cured resin and the film, as shown in Figure~\ref{fig:supporting_soft_stereolithography}.C. While printing introduced slight distortions, they did not affect functionality, as our metamaterial relies on local resonances rather than perfect lattice periodicity.

\newpage
\subsection{Filling Operation}
\label{sec:filling}

\begin{figure}[!h]
\centering
\includegraphics[width=178mm]{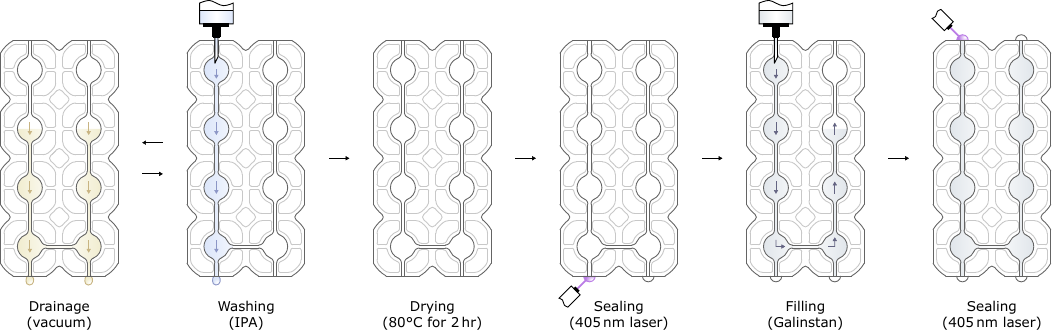}
\caption{Fabrication process for liquid metal pockets within the soft metamaterial, shown from left to right in chronological order. This includes cycles of uncured resin drainage and IPA flushing, followed by oven drying, bottom-side sealing to form U-shaped channels, top-down injection of Galinstan, and final top-sealing via resin droplets that are subsequently UV-hardened.}
\label{fig:supporting_filling}
\end{figure}

The 3D-printed lattice underwent repeated syringe vacuuming and IPA flushing until all uncured residue was removed. This was followed by oven drying at \SI{80}{\celsius} for \SI{2}{\hour} to limit IPA-induced swelling. The channels were selectively sealed with droplets of resin, hardened using a \SI{405}{\nano\meter}-UV laser. This enabled injection of liquid metal without trapping air. This process is illustrated in \textbf{Figure~\ref{fig:supporting_filling}}~and further demonstrated in Movie~S1 using dyed water instead of liquid metal for visualization.

\newpage
\subsection{Sample Fabrication and Mass Properties}
\label{sec:samples}

Using the soft stereolithography and liquid metal filling methods previously described, we manufactured a $4^3$ metamaterial prototype. Its mass properties, along with those of the silicone reference samples, are provided in \textbf{Table~\ref{tab:mass_metamaterial_silicone_samples}}. Additionally, ten standalone unit cells were fabricated using the optimized dimensions $a=\SI{16.5}{\milli\meter}$ and $r=0.42$. Their masses are detailed in \textbf{Table~\ref{tab:mass_standalone_unit_cells}}, where the narrow distribution confirms a highly repeatable manufacturing process.

\begin{table}[!h]
    \centering
    \small
    \renewcommand{\arraystretch}{1.2} 
    \setlength{\tabcolsep}{1.2mm} 
    \caption{Effective mass properties of metamaterial and silicone samples.}
    \label{tab:mass_metamaterial_silicone_samples}
    \begin{threeparttable}
        \begin{tabular}{l S[table-format=1.2] S[table-format=3.1]}
            \toprule
            Sample & {Density (\SI[inter-unit-product = \!\cdot\!]{}{\gram\per\centi\meter\cubed})} & {Mass (\SI{}{\gram})} \\
            \midrule
            \textit{Metamaterial} & & \\
            CAD model (empty) & 0.27 & 76.7 \\
            CAD model (filled) & 0.52 & 150.7 \\
            Prototype (empty) & 0.28 & 80.1 \\
            Prototype (filled) & 0.53 & 151.2 \\
            \addlinespace
            \textit{Plain silicone} & & \\
            Ecoflex 00-10 & 1.03 & 296.3 \\
            Ecoflex 00-20 & 1.01 & 291.6 \\
            Ecoflex 00-30 & 1.02 & 292.5 \\
            \addlinespace
            \textit{Silicone with glass bubbles} & & \\
            Ecoflex 00-10 + K20 & 0.67 & 192.0 \\
            Ecoflex 00-20 + K20 & 0.64 & 182.9 \\
            Ecoflex 00-30 + K20 & 0.65 & 188.1 \\
            \bottomrule
        \end{tabular}
    \end{threeparttable}
\end{table}

\begin{table}[!h]
    \centering
    \small
    \renewcommand{\arraystretch}{1.2} 
    \setlength{\tabcolsep}{1.2mm} 
    \caption{Effective mass properties of standalone unit cells.}
    \label{tab:mass_standalone_unit_cells}
    \begin{threeparttable}
        \begin{tabular}{l @{\hskip 4mm} *{10}{r} @{\hskip 5mm} r}
            \toprule
            Unit cell No. & 1 & 2 & 3 & 4 & 5 & 6 & 7 & 8 & 9 & 10 & Mean $\pm$ SD\\    
            \midrule
            Liquid metal mass (\SI{}{\gram}) & 1.055 & 1.058 & 1.061 & 1.055 & 1.045 & 1.076 & 1.058 & 1.052 & 1.046 & 1.062 & $1.057 \pm 0.009$ \\
            Total mass (\SI{}{\gram})     & 2.066 & 2.064 & 2.058 & 2.059 & 2.056 & 2.075 & 2.057 & 2.053 & 2.043 & 2.045 & $2.058 \pm 0.010$ \\
            \bottomrule
        \end{tabular}
    \end{threeparttable}
\end{table}

\subsection{Vibrating Apparatus}
\label{sec:shaker}

\begin{figure}[!h]
\centering
\includegraphics[width=178mm]{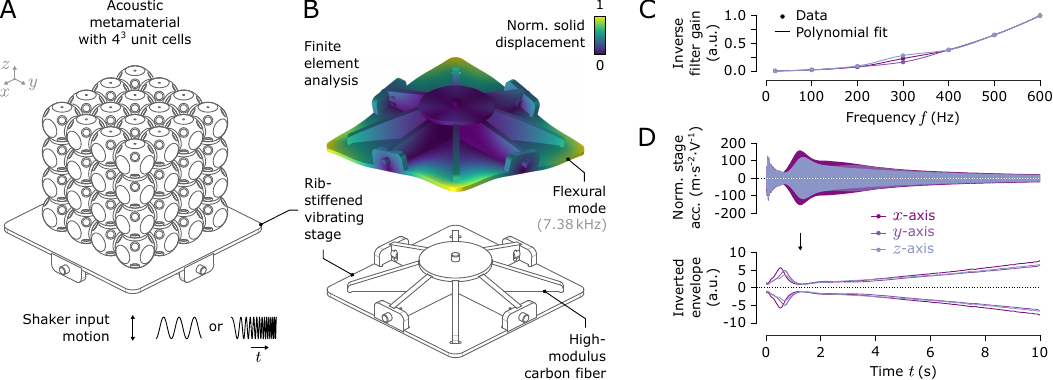}
\caption{A) Rib-stiffened vibrating stage made from high-modulus carbon fiber panels. B) Modal analysis showing the first eigenmode well beyond the target frequency range. C) Inverse filter for optical vibrometry, ensuring constant displacement amplitude at layer $\mathrm{L}_1$. D) Calibration protocol for frequency-independent acceleration. The top graph displays the uncorrected stage response, i.e. acceleration normalized by the analog drive voltage, during a sine sweep. The bottom graph shows the derived inverted voltage envelope used to modulate the input signal. Data corresponds to the filled metamaterial configuration.}
\label{fig:supporting_shaker}
\end{figure}

The vibrating stage supporting the metamaterial was designed to avoid parasitic modes and strong nonlinearities within the \SI{0}{\hertz} to \SI{600}{\hertz} range. It was constructed using high-modulus carbon fiber panels (DragonPlate, Allred \& Associates). As illustrated in \textbf{Figure~\ref{fig:supporting_shaker}}.A, they were assembled in a rib-stiffened configuration and bonded with structural epoxy (2216, 3M Company) to create a lightweight, rigid structure. Its modal response was computed via finite element analysis in COMSOL Multiphysics 6.2. Based on supplier specifications, the composite material was modeled as isotropic, homogeneous, and linear elastic, with a Poisson's ratio $\nu=0.3$, a density $\rho_s=\SI[inter-unit-product = \!\cdot\!]{1.61}{\gram\per\centi\meter\cubed}$, and a Young's modulus $E=\SI{105}{\giga\pascal}$. Simulation results in Figure~\ref{fig:supporting_shaker}.B identify the first eigenmode at \SI{7.38}{\kilo\hertz}. This is more than an order of magnitude above the target range, thus ensuring that stage dynamics did not couple with that of the metamaterial.

Although our stage design prevents structural resonances, the electrodynamic shaker exhibits an intrinsic non-uniform frequency response that must be compensated for. To ensure frequency-independent excitation, we employed two distinct calibration protocols. For the optical vibrometry presented in Figures~3~and~4, the driving voltage was iteratively tuned to maintain a constant displacement amplitude at the first metamaterial layer, $\mathrm{L}_1$. As shown in Figure~\ref{fig:supporting_shaker}.C, measurements at six discrete frequencies between \SI{0}{\hertz}~and~\SI{600}{\hertz} were used to construct an inverse filter, defined by a piecewise cubic polynomial fit. Conversely, for the acceleration analysis in Figure~5, we performed a \SI{10}{\second} sine sweep over the target bandwidth to generate a look-up table from the inverted acceleration envelope of the stage, as shown in Figure~\ref{fig:supporting_shaker}.D. This approach maintained a constant acceleration amplitude, set to \SI[inter-unit-product = \!\cdot\!]{13}{\meter\per\second\squared} to maximize signal-to-noise ratio while preventing saturation of the~shaker. In both instances, the lack of sharp spectral features in the calibration profiles confirmed that the platform remained free of parasitic modes.

\subsection{Signal Conditioning Electronics}
\label{sec:electronics}

\begin{figure}[!h]
\centering
\includegraphics[width=178mm]{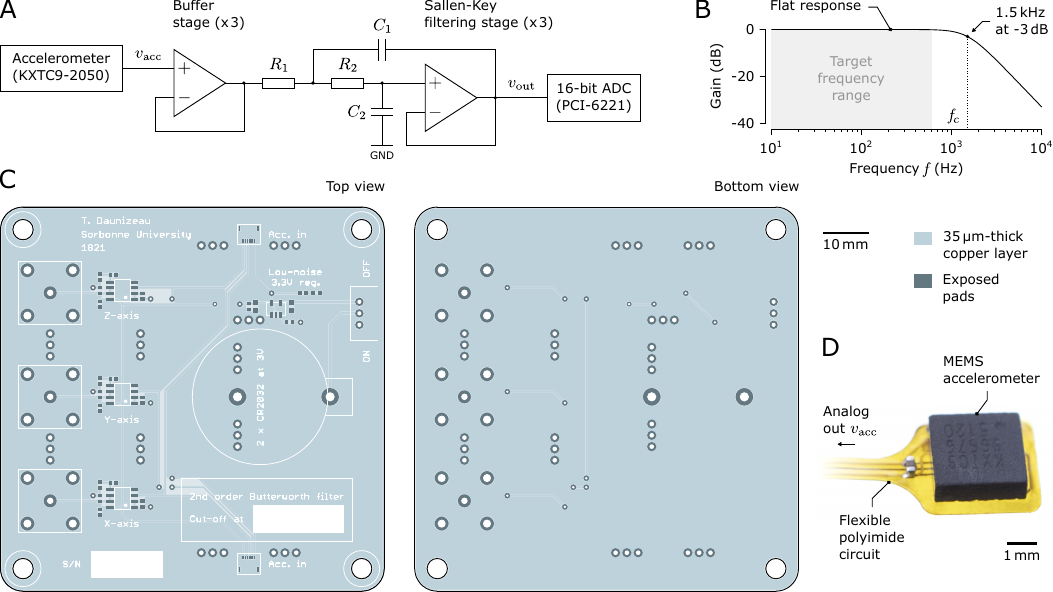}
\caption{A) Circuit schematic of the analog Sallen-Key anti-aliasing filter. B) Corresponding frequency response of the \SI{1.5}{\kilo\hertz} second-order Butterworth filter, showing a maximally flat passband. C) Top and bottom PCB layouts of the~battery-powered, low-noise signal conditioning boards. D) MEMS accelerometer mounted on a bespoke \SI{100}{\micro\meter}-thick polyimide substrate.}
\label{fig:supporting_electronics}
\end{figure}

Both the top (metamaterial) and bottom (vibrating stage) MEMS accelerometers were~mounted onto custom flexible circuits fabricated from a \SI{100}{\micro\meter}-thick polyimide substrate (Kapton, DuPont), as depicted in \textbf{Figure~\ref{fig:supporting_electronics}}.D. Their low mass of \SI{40}{\milli\gram} and high compliance ensured they did not alter the metamaterial dynamics. These flexible circuits were connected to signal conditioning boards where the analog acceleration signals, $v_\mathrm{acc}(t)$, were fed into a preliminary buffer stage, followed by an active anti-aliasing filter using a Sallen-Key topology, as shown in Figure~\ref{fig:supporting_electronics}.A. This produced the output voltage, $v_\mathrm{out}(t)$, following the unity-gain transfer function, $V_{\!\mathrm{out}}(s)/V_{\!\mathrm{acc}}(s)$, expressed as
\begin{equation}\label{eq:transfer_function_sallen_key}
\frac{V_{\!\mathrm{out}}(s)}{V_{\!\mathrm{acc}}(s)} = \frac{1}{C_1 C_2 R_1 R_2 \, s^2 + C_2 (R_1\! +\! R_2) \, s + 1}
\, ,
\end{equation}
where $s$ is the Laplace variable, and $C_1$, $C_2$, $R_1$, and $R_2$ are the capacitor and resistor values,~respectively. The cut-off frequency, $f_{\:\!\!c}$, was defined as
\begin{equation}\label{eq:cutoff_sallen_key}
f_{\:\!\!c} = \frac{1}{2\pi\sqrt{C_1 C_2 R_1 R_2}} 
\, ,
\end{equation}
and the damping ratio, $\zeta$, was given by
\begin{equation}\label{eq:zeta_sallen_key}
\zeta = \frac{C_2 (R_1\! +\! R_2)}{2\sqrt{C_1 C_2 R_1 R_2}} 
\, .
\end{equation}
To emulate a second-order Butterworth low-pass filter, the damping ratio $\zeta$ was tuned to $\sqrt{2}/2$.~Setting the cut-off frequency to \SI{1.5}{\kilo\hertz} ensured a maximally flat response over the \SI{0}{\hertz} to \SI{600}{\hertz} operating range, as shown in Figure~\ref{fig:supporting_electronics}.B. This was achieved with precision, low-ppm passives, $C_1=\SI{10}{\nano\farad}$, $C_2=\SI{4.7}{\nano\farad}$, $R_1=\SI{20}{\kilo\ohm}$, and $R_2=\SI{12}{\kilo\ohm}$. To enhance signal integrity, the purpose-built conditioning boards were battery-powered and featured a low-noise voltage regulator (LT1761, Linear Technology), with the board layout detailed in Figure~\ref{fig:supporting_electronics}.C.

\subsection{Image Processing}
\label{sec:image_processing}

\begin{figure}[!t]
\centering
\includegraphics[width=178mm]{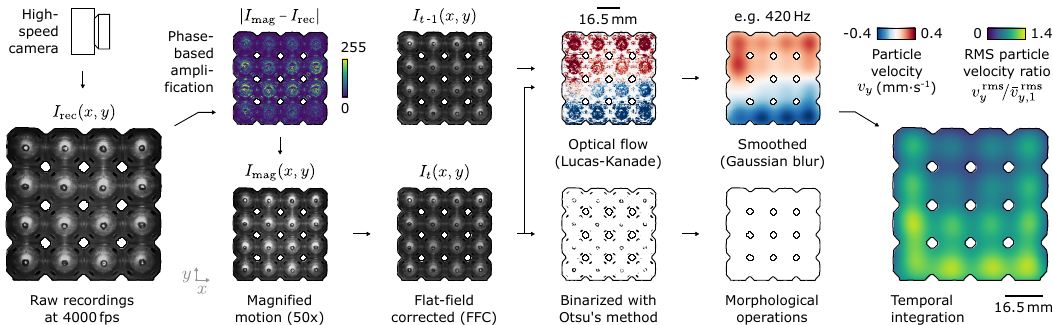}
\caption{Processing workflow for full-field vibrometry of the metamaterial, exemplified by a \SI{420}{\hertz} sine input along the $y$-axis. The pipeline involves high-speed imaging, illumination correction, phase-based motion amplification, and optical flow estimation. The resulting sparse particle velocity field is smoothed via Gaussian blur, and regions of interest are isolated using binary masks. The RMS field is derived by temporal integration.}
\label{fig:supporting_image_processing}
\end{figure}

Comprehensive details regarding the image processing workflow used to capture the particle velocity field $\boldsymbol{v}$ are provided in the Experimental Section. The individual stages are illustrated in \textbf{Figure~\ref{fig:supporting_image_processing}} for a \SI{420}{\hertz} excitation along $\boldsymbol{e}_y$, a frequency exceeding the band gap $\mathrm{BG}_y$. The motion recorded in the raw frames, $I_\mathrm{rec}(x,y)$, was first magnified fiftyfold to yield $I_\mathrm{mag}(x,y)$. Following illumination correction, the frame at time $t$, denoted $I_t(x,y)$, was compared with the preceding frame $I_{t\,\minus 1}(x,y)$ using optical flow to derive $\boldsymbol{v}$. As the raw flow data is sparse, i.e. valid only at successfully tracked features, a Gaussian blur was applied to generate a spatially continuous field. Finally, binary masks generated via Otsu's binarization and refined by morphological operations were applied to isolate the metamaterial structure. In Figure~\ref{fig:supporting_image_processing}, these masks are applied at every step for visual clarity.

\newpage\section{Supporting Movies}
\label{sec:movies}

\subsection{Movie S1}
This movie illustrates the filling of the U-shaped mesofluidic channels using colored fluid and a metamaterial slice, as in Figure~1.C. It further demonstrates structural compliance.

\subsection{Movie S2}
This video presents the finite element model along with its underlying assumptions for solid and liquid phases. It further shows the flexural and torsional resonances driving band gap formation.

\subsection{Movie S3}
This movie presents the experimental frequency response of individual unit cells, captured via a see-through imaging technique that reveals the out-of-phase motion of the liquid metal pocket.

\subsection{Movie S4}
This movie presents an experimental vibrometry analysis of our soft 3D metamaterial via high-speed imaging, detailing both the image processing workflow and the resulting band gap data.

\bibliographystyle{MSP}
\bibliography{biblio_soft_metamaterial}
